\PassOptionsToPackage{table}{xcolor}
\documentclass[a4paper,12pt]{article}
\pdfoutput=1 

\usepackage{jheppub} 

\usepackage[T1]{fontenc} 

	\usepackage{amsthm}
	\usepackage{tikz}
		\usetikzlibrary{calc}
\usetikzlibrary{arrows,positioning} 
\usetikzlibrary{automata, arrows.meta}
\usepackage[percent]{overpic}
\usepackage{slashed}
\usepackage{wrapfig}
\usepackage{tabu}
\usepackage{diagbox}
\usepackage{mathrsfs,amsmath,amssymb,amsthm,amsfonts,tikz,graphicx,accents,hyperref, color}
\usepackage{dsfont,epiolmec, latexsym, stmaryrd, comment}
\usepackage{slashed,ccaption}
\usepackage{mathrsfs, calligra}
\usepackage{leftidx}
\usepackage{import}
\usepackage{multirow}
\usepackage{amsfonts}
\usepackage{pifont}
\usepackage{tabularx}
\usepackage[utf8]{inputenc}
\usetikzlibrary{intersections,calc}
\usepackage{tikz-3dplot}
\usepackage{ifthen}
\usetikzlibrary{arrows}
\usepackage{amsmath}
\usepackage{xcolor}
\usepackage{geometry}

\tikzset{
    >=stealth',
    punkt/.style={
           rectangle,
           rounded corners,
           draw=black, very thick,
           text width=6.5em,
           minimum height=2em,
           text centered},
    pil/.style={
           ->,
           thick,
           shorten <=2pt,
           shorten >=2pt,}
}

\usepackage{array}
\DeclareFontFamily{OT1}{rsfs}{}

\DeclareFontShape{OT1}{rsfs}{m}{n}{ <-7> rsfs5 <7-10> rsfs7 <10->rsfs10}{} 

\DeclareMathAlphabet{\mycal}{OT1}{rsfs}{m}{n}

\newcommand{\thh}{{$\mathfrak{H}_{3}$}}

\newcommand{\eps}{\varepsilon}

\newcommand{\be}{\begin{equation}}
\newcommand{\ee}{\end{equation}}

\tdplotsetmaincoords{48}{112}
\definecolor{darkgreen}{rgb}{0.1,0.6,0.1}
\definecolor{darkblue}{rgb}{0,0,0.3}
\definecolor{darkred}{rgb}{0.7,0,0}

\definecolor{light gray}{RGB}{220,220,220}
\definecolor{dark purple}{RGB}{108,0,217}
\definecolor{pink}{RGB}{190,20,100}
\definecolor{orang}{RGB}{193,63,0}
\definecolor{green}{RGB}{11,98,17}
\definecolor{darkpink}{RGB}{153,0,76}
\definecolor{bluegreen}{RGB}{0,102,102}
\definecolor{greenlagan}{RGB}{0,102,0}
\definecolor{redgreen}{RGB}{102,102,0}
\definecolor{Redgreen}{RGB}{153,76,0}
\definecolor{vividviolet}{rgb}{0.62, 0.0, 1.0}
\definecolor{amaranth}{rgb}{0.9, 0.17, 0.31}
\definecolor{palatinateblue}{rgb}{0.15, 0.23, 0.89}
\definecolor{brightpink}{rgb}{1.0, 0.0, 0.5}
\definecolor{cornflowerblue}{rgb}{0.39, 0.58, 0.93}
\definecolor{deepcarminepink}{rgb}{0.94, 0.19, 0.22}
\definecolor{radicalred}{rgb}{1.0, 0.21, 0.37}

\makeatletter \@addtoreset{equation}{section}

\newcommand\hnote[1]{\textcolor{magenta}{\bf [Hamid:\,#1]}}

\newcommand\mnote[1]{\textcolor{red}{\bf [Martin:\,#1]}}

\title{{\Large{Boundary Heisenberg Algebras and Their Deformations}}}
\author[a]{Mart\'in Enr\'iquez Rojo}
\author[b]{and H. R. Safari}

\affiliation{$^a$ Arnold Sommerfeld Center for Theoretical Physics, Ludwig-Maximilians-Universit\"at \\
Theresienstra\ss e 37, 80333 M\"unchen, Germany}
\affiliation{$^b$ School of Physics, Institute for Research in Fundamental
Sciences (IPM),\\ P.O.Box 19395-5531, Tehran, Iran}

\emailAdd{martin.enriquez@physik.lmu.de}
\emailAdd{hrsafari@ipm.ir}

\abstract{
We investigate the deformations and rigidity of boundary Heisenberg-like algebras. In particular, we focus on the Heisenberg and $\text{Heisenberg}\oplus\mathfrak{witt}$ algebras which arise as symmetry algebras in three-dimensional gravity theories. As a result of the deformation procedure we find a large class of algebras. While some of these algebras are new, some of them have already been obtained as asymptotic and boundary symmetry algebras, supporting the idea that symmetry algebras associated to diverse boundary conditions and spacetime loci are algebraically interconnected through deformation of algebras. The deformation/contraction relationships between the new algebras are investigated. In addition, it is also shown that the deformation procedure reaches new algebras inaccessible to the Sugawara construction. As a byproduct of our analysis, we obtain that $\text{Heisenberg}\oplus\mathfrak{witt}$ and the asymptotic symmetry algebra Weyl-$\mathfrak{bms}_3$ are not connected via single deformation but in a more subtle way.}

\begin{document}
\begin{flushright}
  {\small 
  LMU-ASC 46/21 \\
   IPM/P-2021/40
  }
\end{flushright}
\maketitle
\section{Introduction and motivation}


In the recent years,  pioneering analysis of asymptotically flat spacetimes performed by Bondi, van der Burg, Metzner and Sachs (BMS) \cite{Bondi:1962px,Sachs:1962wk,Sachs:1962zza} has been refined and extended to many other dimensions, spacetimes and boundaries.
Notably, the structure of spacetime near generic null surfaces (including event and cosmological horizons), not only but mostly in three spacetime dimensions, has been intensively investigated in the last few years \cite{Afshar:2016kjj,Afshar:2016uax, Afshar:2016wfy,Grumiller:2019fmp}. In this context, boundary Heisenberg algebras have played a major role all the way through and constitute a fundamental piece behind the different symmetry algebras popping up in a variety of boundary symmetry analysis. 

Contemporaneously and linked to these developments, a new research area exploring deformations of these infinite dimensional symmetry algebras has emerged \cite{FarahmandParsa:2018ojt, Safari:2019zmc, Safari:2020pje}. A major reason behind it is that different symmetry algebras arise from different boundary conditions imposed at the same loci. From a physical viewpoint, it would be desirable to have a better understanding of this. Several attempts have been carried out in this regard. For example, a thermodynamical interpretation for the different boundary conditions has been pursued in \cite{Adami:2021kvx}, the idea of connecting different symmetry algebras through changes of slicing and the existence of a fundamental slicing, where the null boundary symmetry algebra is $\text{Heisenberg}\oplus\text{Diff}(d-2)$, have been proposed and explored in \cite{Adami:2020ugu,Adami:2021nnf}, and several claims on the closure and lack of central extensions of these algebras have been made in \cite{Freidel:2021cjp,Ciambelli:2021vnn}. Nevertheless, another possible approach is that boundary algebras associated to different boundary conditions should be connected via deformations, constituting families of deformations which might unveil and help to discern properties coming from possible diverse choices of boundary conditions. Ultimately, it is expected that the deformation analysis of the corresponding algebras can shed light on how to better select and understand choices of boundary conditions. Examples validating such an approach can be found in \cite{Safari:2020pje} and new ones will be encountered along this work.

In addition, different symmetry algebras come from the analysis at different boundaries. These are also expected to be related in a unique framework. For example, asymptotic and near horizon symmetry algebras have been interpolated in \cite{Grumiller:2019ygj}. The deformation analysis has also been fruitful in relating such algebras. In fact, it has been shown that certain near horizon and asymptotic symmetry algebras of three-dimensional and four-dimensional asymptotically flat and Friedman spacetimes form part of the same multi-parametric families of deformation algebras, denoted as $W$-algebras\footnote{Not to be confused with $\mathcal{W}$-algebras present in the context of conformal field theories \cite{Zamolodchikov:1985wn}.}  \cite{Safari:2019zmc,Safari:2020pje,Enriquez-Rojo:2021blc,Enriquez-Rojo:2021rtv}.

Another motivation for studying deformations of infinite dimensional algebras is that they provide us with a path to construct new algebras, which can possibly be realized as symmetry algebras under new boundary conditions and in different geometric settings. In fact, some examples of this procedure will be illustrated along this work.

Bearing this in mind, we explore  deformations of boundary Heisenberg algebras in this work. For technical simplicity, we restrict ourselves to the three-dimensional case, although we expect the main features of this analysis to follow in higher dimensions. In particular, we explore deformations of the infinite dimensional Heisenberg algebra \eqref{Dirac-3d} and $\text{Heisenberg}\oplus\mathfrak{witt}$ \eqref{Heisenberg-direct-sum-algebra-3d}. Our main results can be summarized as follows:
\begin{enumerate}
    \item We have obtained linear, formal and infinitesimal deformations of the Heisenberg algebra \eqref{Dirac-3d}. Among the new algebras, we find several well-known asymptotic and near horizon symmetry algebras, supporting the previous discussion on the role that deformations play in interpolating between different symmetry algebras. Two three-parametric deformation  families of algebras, ``mother algebras'', arise in this analysis, $\widehat{W}_{\nu}(0,b)$ \eqref{Deform-G,F,BarG} and $\mathcal{H}_3(\alpha,\nu,\eta)$ \eqref{Deform-G,F,BarF,BarG,TildeF,TildeG}, together with their respective central extensions. Selected new algebras, as well as their relations through deformation and contraction, are summarized in figure \ref{Fig.1}. 
    \item We have shown that, although some of the new algebras obtained through deformation can be also found by means of Sugawara constructions, this does not hold true in general. This points towards the fact that the deformation procedure yields more algebras. Concretely, we find evidence that those algebras belonging to the family $\widehat{W}_{\nu}(0,b)$ cannot be obtained via Sugawara construction for arbitrary $b$, while those within $\mathcal{H}_3(\alpha,\nu,\eta)$ can be reached using this procedure.
    \item We derive linear, formal and infinitesimal deformations of the $\text{Heisenberg}\oplus\mathfrak{witt}$ algebra \eqref{Heisenberg-direct-sum-algebra-3d} with and without allowing for central extensions. Interestingly, all the deformations we obtained come uniquely from deforming the Heisenberg part of the algebra. Among the new deformations, we have found the direct sum of three Virasoro algebras or the direct sum of centrally extended $W(0,b)$ and the Virasoro algebra which have been realized as asymptotic symmetry algebras of Maxwell Chern-Simons and AdS-Lorentz Chern-Simons gravity theories in \cite{Adami:2020xkm,Concha:2018jjj}. Furthermore, we also obtained the algebra $\mathfrak{vir}\oplus\mathfrak{vir}\oplus\mathfrak{u}(1)$ which has been identified as asymptotic symmetry algebra of AdS$_3$ when extra Weyl symmetry is considered \cite{Alessio:2020ioh,Geiller:2021vpg}.
    \item This analysis served us to rule out the possibility that this near horizon algebra could be related through direct deformation to the Weyl-BMS algebra \eqref{eq:weylbms3} and to the null boundary algebra given by equation (3.34) in \cite{Adami:2020ugu}. Instead, we showed that the relation with \eqref{eq:weylbms3} is more subtle and consists of a contraction followed by a double deformation procedure. Alternatively, one can start from a twisted double Heisenberg algebra \eqref{eq:toWeyl-BMS1} and reach $\text{Heisenberg}\oplus\mathfrak{witt}$, as well as Weyl-BMS and equation (3.34) in \cite{Adami:2020ugu}.
\end{enumerate}


This paper is organized as follows: in section \ref{sect_review}, we briefly list the boundary and asymptotic symmetry algebras directly involved in our analysis. In sections \ref{sec_defsHeisenbeg} and \ref{sect_weylbms}, we respectively consider infinitesimal and formal deformations of the infinite dimensional Heisenberg \eqref{Dirac-3d} and $\text{Heisenberg}\oplus\mathfrak{witt}$  \eqref{Heisenberg-direct-sum-algebra-3d} algebras. Our conclusions and further remarks are contained in section \ref{sec-discussions}. Taking into account that the main physical motivation for considering Heisenberg-like algebras comes from their role in the spacetime structure near generic null surfaces, we review how these algebras are derived in appendix \ref{app_spacetimeGNS}. Appendix \ref{app_deformations} contains a review of deformation theory and its relation to cohomology of Lie algebras. Therein we describe the deformations we investigate and the methodology we follow in this text. The computational details concerning the analysis of Jacobi identities, necessary to obtain the allowed deformations, and relevant study of the inverse procedure of deformations, the so-called contractions, are relegated to the appendices \ref{app_Jacobis} and \ref{app_Heisenbergfromcontractions}. Besides, fearing the risk that the reader could miss important details and results in the middle of an extensive technical treatment, we include remarks with interesting results and details all along the text.

\paragraph{Notation.} We use generally ``mathfrak'' font for the algebras, e.g. $\mathfrak{vir}, \mathfrak{witt}$ and $\mathfrak{H}_3$ for the Virasoro, Witt and infinite dimensional Heisenberg algebras respectively. Their centrally extended versions will be denoted by a hat, e.g. $\mathfrak{vir}=\widehat{\mathfrak{witt}}$. We will use sub-indices for distinguished deformations of main algebras, e.g. $\mathfrak{H}_{3\nu\alpha}$ corresponds to a two-parameter ($\nu$ and $\alpha$) deformation of $\mathfrak{H}_3$. We will be using ``$W(a,b)$ family'' of algebras to denote a set of algebras for different values of the $a,b$ parameters.

\section{Review of symmetry algebras in three spacetime dimensions}
\label{sect_review}
In this section, we collect relevant three-dimensional near horizon and asymptotic symmetry algebras for our work. 

\subsection{Near horizon symmetry algebras}

Let us briefly list the near horizon symmetry algebras which will play a major role in our analysis. For more details on the derivation of these and more boundary symmetry algebras in any spacetime dimension, we refer the reader to appendix \ref{app_spacetimeGNS}.

\paragraph{Heisenberg $\oplus$ Diff($S^1$).} 
In three spacetime dimensions, the algebra of charges in the ``Heisenberg-Direct sum slicing'' \eqref{Heisenberg-direct-sum-algebra}, also called ``fundamental slicing'', takes the form \cite{Adami:2020ugu}
\begin{subequations}\label{Heisenberg-direct-sum-algebra-3d}
    \begin{align}
        &[{\cal Q}_m,{\cal Q}_n]=[\mathcal{P}_m,\mathcal{P}_n]=0 \ ,\\
        &[{\cal Q}_m,\mathcal{P}_n]= i\hbar \delta_{m+n,0} \ ,\\
        &[\mathcal{J}_m,{\cal Q}_n]=[\mathcal{J}_m,\mathcal{P}_n]=0 \ ,\\
        &[\mathcal{J}_m,\mathcal{J}_n]= (m-n)\mathcal{J}_{m+n} +\frac{c}{12}m^3 \delta_{m+n,0} \ ,
    \end{align}
\end{subequations}
where we have also added a central term $c$ which arises in the TMG case \cite{Adami:2021sko}.

\paragraph{Virasoro-Kac-Moody algebra.} The ``non-expanding null surface algebra'' \eqref{BMS-like-non-expanding-algebra} takes the following form in three spacetime dimensions
\begin{subequations}\label{BMS-like-non-expanding-algebra-3d}
    \begin{align}
         &[{\cal Q}_m,{\cal Q}_n]={\tilde{c}}\,m\, \delta_{m+n,0} \ ,\\
        &[\mathcal{J}_m,{\cal Q}_n]=- n{\cal Q}_{m+n}+{\bar{c}}\,m^2\, \delta_{m+n,0} \ ,\\
        &[\mathcal{J}_m,\mathcal{J}_n]= (m-n)\mathcal{J}_{m+n}+\frac{c}{12}\,m^3 \delta_{m+n,0} 
    \end{align}
\end{subequations}
after the inclusion of central extensions. This algebra is the member $\widehat{W}(0,0)$ of a wider family of near horizon symmetry algebras $\widehat{W}(0,b)$  \cite{FarahmandParsa:2018ojt,Grumiller:2019fmp}, which will be described in the next subsection. Furthermore, this algebra can be also realized as asymptotic symmetry algebra in $(A)dS_3$ by imposing Compère-Song-Strominger (CSS) boundary conditions \cite{Compere:2013bya} or as a near-horizon symmetry algebra in \cite{Afshar:2015wjm}. In the latter cases, the $\mathcal{Q}$ generators are interpreted as supertranslations, while in appendix \ref{app_spacetimeGNS} they correspond to Weyl generators. Besides, it is worth to point out that only $c$ and $\tilde{c}$ show up as a central extension in \cite{Compere:2013bya}, while the three central terms arise in \cite{Afshar:2015wjm}. 

\paragraph{Heisenberg-like algebra.} In three dimensions, the ``non-expanding null surface algebra in Heisenberg slicing'' \eqref{Poisson-CR1} turns out to be
\begin{equation}\label{Dirac-3d}
    \begin{split}
    &[\mathcal{J}_{m},\mathcal{J}_{n}]=0 \ ,\\
    &[\mathcal{J}_{m},\mathcal{P}_{n}]=i \hbar \,m\delta_{m+n,0} \ ,\\
        &[\mathcal{P}_{m},\mathcal{P}_{n}]=0 \ .
    \end{split}
\end{equation}
By means of a redefinition $\mathcal{J}_{m}\rightarrow m\,\mathcal{J}_{m}$, one gets the infinite dimensional Heisenberg algebra \cite{Afshar:2016kjj, Afshar:2016wfy}, denoted by $\mathfrak{H}_3$ \footnote{For the sake of clarity and computational simplicity, we rescale the generators such that the factor $i\hbar\to1$, leading to $[\mathcal{J}_{m},\mathcal{P}_{n}]=m\delta_{m+n,0}$.}. It is worth pointing out that this algebra can also be obtained in a two-dimensional analysis as discussed in \cite{Adami:2020ugu}.

In this work, we analyze deformations and stability of the infinite dimensional algebras \eqref{Dirac-3d} and \eqref{Heisenberg-direct-sum-algebra-3d} which involve a Heisenberg part \footnote{Let us point out that the rescaling $\mathcal{J}_{m}\rightarrow m\,\mathcal{J}_{m}$ affects the zero mode of the infinite dimensional Heisenberg algebra. We consider deformations of \eqref{Dirac-3d}, which does not include the zero mode, along this work. Nonetheless, we have noticed that there are significant changes if we instead deform $[\mathcal{J}_{m},\mathcal{P}_{n}]=\delta_{m+n,0}$. Indeed, the new algebras reached via deformation include \eqref{MostDeform-JJ}, \eqref{Deform-JP-barG} for $\beta=0$, \eqref{Deform-JP-barF} for $\eta=0$, \eqref{Deform-G,F,BarG-3} and \eqref{Deform-G,BarG}, all of them without the coefficient $m$ in the second commutator. Nevertheless, the deformation \eqref{Deform-F,BarG}, without the coefficient $m$ in the second commutator, is only allowed for $b=1$, which is a very important difference. An exhaustive treatment of such deformations is beyond the scope of this work, although we highly encourage it.}.  



\subsection{Asymptotic symmetry algebras}

During the last years and inspired by the earlier work of Bondi, van der Burg, Metzner and Sachs (BMS) \cite{Bondi:1962px,Sachs:1962wk,Sachs:1962zza}, the infrared structure of several spacetimes has been revisited and further explored. As a consequence, many new asymptotic symmetry algebras of diffeomorphisms and charges have arisen in different spacetime dimensions and diverse background spacetimes (prominently including flat \cite{Bondi:1962px,Sachs:1962wk,Sachs:1962zza,Barnich:2009se,Barnich:2010eb,Barnich:2011mi,Campiglia:2014yka,Campiglia:2015yka,Freidel:2021fxf}, (Anti-) de Sitter \cite{Brown:1986nw,Compere:2015knw,Compere:2019bua,Compere:2020lrt} and FLRW \cite{Bonga:2020fhx,Enriquez-Rojo:2020miw,Enriquez-Rojo:2021blc}). We do not aim here for an exhaustive review but we shall only mention those three-dimensional cases which will be connected to our deformation analysis, besides the well-known $\mathfrak{witt}$ and $\mathfrak{vir}$ algebras coming from standard analysis of asymptotically $(A)dS_3$ spacetimes \footnote{To be more accurate, by imposing the Brown-Henneaux boundary condition \cite{Brown:1986nw} in $(A)dS_3$, one obtains two copies of Virasoro algebras $\mathfrak{vir}\oplus\mathfrak{vir}$.}.

\paragraph{$\mathfrak{bms}_3$ and $W(a,b)$. }

The algebra $\mathfrak{bms}_3$ arises from the study of asymptotically flat spacetimes in three dimensions when the ``supertranslation'' and  ``superrotation'' sectors are present but Weyl scalings are not allowed \cite{Barnich:2010eb}. $\mathfrak{bms}_3$ admits a two-parametric family of deformations \cite{FarahmandParsa:2018ojt,Safari:2020pje} denoted as $W(a,b)$ and given by:
\begin{subequations}\label{eq:Wab}
\begin{align}
    & [\mathcal{J}_m,\mathcal{J}_n]=(m-n)\mathcal{J}_{m+n} \ , \\
    & [\mathcal{J}_m,\mathcal{P}_{n}]=-(n+bm+a)\mathcal{P}_{m+n} \ , \\
    & [\mathcal{P}_m,\mathcal{P}_{n}]=0 \ ,
\end{align}
\end{subequations}
where $\mathcal{J}_m$ and $\mathcal{P}_m$ represent, respectively, the superrotation and supertranslation generators and $\mathfrak{bms}_3\simeq W(0,-1)$. The $W(0,b)$ algebras are also realized as near-horizon symmetry algebras of 3-dimensional black holes \cite{Grumiller:2019fmp}. Concretely, the Virasoro-Kac-Moody algebra \eqref{BMS-like-non-expanding-algebra-3d} corresponds to $W(0,0)$ and can also be obtained as asymptotic symmetry algebra in $AdS_3$ by imposing CSS boundary conditions \cite{Compere:2013bya}. Central extensions of both algebras are denoted by $\widehat{\mathfrak{bms}}_3$ and $\widehat{W}(a,b)$.

\paragraph{Weyl-BMS.} It was recently pointed out that, besides ``supertranslations'' and  ``superrotations'', one can admit also Weyl/scaling transformations \cite{Adami:2020ugu,Donnay:2020fof,Batlle:2020hia,Freidel:2021fxf}. In the three-dimensional case, the algebra of asymptotically flat spacetimes, called 3D Weyl-BMS and denoted as $\mathfrak{bmsw}_3$, is augmented to
\begin{subequations}\label{eq:weylbms3}
\begin{align}
    & [\mathcal{J}_m,\mathcal{J}_n]=(m-n)\mathcal{J}_{m+n} \ , \\
    & [\mathcal{J}_m,\mathcal{P}_{n}]=(m-n)\mathcal{P}_{m+n} \ , \\
   & [\mathcal{J}_m,\mathcal{D}_{n}]=-n\mathcal{D}_{m+n} \ , \\
    & [\mathcal{P}_m,\mathcal{P}_{n}]=0 \ , \\
    & [\mathcal{D}_m,\mathcal{P}_n]=\mathcal{P}_{m+n} \ , \\
    & [\mathcal{D}_m,\mathcal{D}_{n}]=0 \ ,
\end{align}
\end{subequations}
where $\mathcal{D}_m$ are the Weyl generators. This algebra admits non-trivial central as well as non-central extensions which can be found in \cite{Fuentealba:2020zkf}. It is worth highlighting that, although at first Weyl symmetry leads to get $\mathfrak{bmsw}_3$ algebra, this is not the only way to realize it. In fact, it has been shown that the Weyl symmetry leads to a larger symmetry algebra $\mathfrak{vir}\oplus\mathfrak{vir}\oplus\mathfrak{u}(1)$ \cite{Alessio:2020ioh, Geiller:2021vpg}.

At last, we would like to point out that $\widehat{\mathfrak{bms}}_3\oplus\mathfrak{vir}$ and $\mathfrak{vir}\oplus\mathfrak{vir}\oplus\mathfrak{vir}$ have been realized as the asymptotic symmetry algebras of 3-dimensional Maxwell Chern-Simons gravity theories invariant under $\mathfrak{iso}(2)\oplus\mathfrak{sl}(2,\mathbb{R})$ and $\mathfrak{so}(2,2)\oplus\mathfrak{sl}(2,\mathbb{R})$ respectively by considering certain boundary conditions \cite{Adami:2020xkm,Concha:2018jjj}.


\section{Deformations of the Heisenberg algebra}\label{sec_defsHeisenbeg}
In this section, we consider infinitesimal and formal deformations of the infinite dimensional Heisenberg algebra $\mathfrak{H}_{3}$ \eqref{Dirac-3d} as described in appendix \ref{app_deformations}. First, we investigate deformations of each commutator separately and then we study the deformations of \thh\, in all commutators at once. In this analysis, we rescale the generators to set $i\hbar\to1$ for convenience.

\subsection{Deformation of separate commutators }
\label{sect_H3separate}

In the following, we deform individually each of the three commutators present in \eqref{Dirac-3d} and find several new algebras as a byproduct.

\subsubsection*{ $\bullet$ Deformations of \texorpdfstring{$[\mathcal{J},\mathcal{J}]$}{JJ}}
We proceed to analyze the possible deformations of two $\mathcal{J}$s 
\begin{equation}\label{Deform-JJ}
    \begin{split}
    &[\mathcal{J}_{m},\mathcal{J}_{n}]=(m-n)F(m,n)\mathcal{J}_{m+n}+(m-n)G(m,n)\mathcal{P}_{m+n}+(m-n)A(m,n) \ ,\\
    &[\mathcal{J}_{m},\mathcal{P}_{n}]=m\delta_{m+n,0} \ ,\\
        &[\mathcal{P}_{m},\mathcal{P}_{n}]=0 \ ,
    \end{split}
\end{equation}
where $F(m,n)$, $G(m,n)$ and $A(m,n)$ are symmetric functions. The Jacobi identities $[\mathcal{J}_{m},[\mathcal{J}_{n},\mathcal{J}_{l}]]+\text{cyclic permutations}=0$ lead to independent relations as
\begin{multline}\label{F-JJ}
(n-l)(m-n-l)F(n,l)F(m,n+l)+(l-m)(n-l-m)F(l,m)F(n,l+m)+ \\
(m-n)(l-m-n)F(m,n)F(l,m+n)=0 \ ,
\end{multline}
\begin{multline}\label{FG-JJ}
(n-l)(m-n-l)F(n,l)G(m,n+l)+(l-m)(n-l-m)F(l,m)G(n,l+m)+ \\
(m-n)(l-m-n)F(m,n)G(l,m+n)=0 \ ,
\end{multline}
and
\begin{multline}\label{FAG-JJ1}
\{(n-l)(m-n-l)F(n,l)A(m,n+l)+(l-m)(n-l-m)F(l,m)A(n,l+m)\\
+(m-n)(l-m-n)F(m,n)A(l,m+n)\}\\
+\{m(n-l)G(n,l)+n(l-m)G(l,m)+l(m-n)G(m,n)\}\delta_{m+n+l,0}=0 \ .
\end{multline}
We now consider the infinitesimal deformations which correspond to new relations including only the first order of functions as
\begin{equation}
    (m(n-l)G(n,l)+n(l-m)G(l,m)+l(m-n)G(m,n))\delta_{m+n+l,0}=0,
\end{equation}
which is solved by $G(m,n)=constant$. \footnote{We would like to note that $G(m,n)=G(n,m)$, what makes $(m(n-l)G(n,l)+n(l-m)G(l,m)+l(m-n)G(m,n))$ antisymmetric under the exchange of either of $m$, $n$ and $l$, as well as symmetric under the simultaneous exchange $m\rightarrow n$, $n\rightarrow l$ and $l\rightarrow m$. Using this information, we could not find a general argument against the presence of other polynomial solutions. Nevertheless, we discarded linear and quadratic ansatz for $G(m,n)$ by explicit computation and firmly believe that this equation has no other solution than $G(m,n)=constant$.}
 


The Jacobi identities $[\mathcal{J}_{m},[\mathcal{J}_{n},\mathcal{P}_{l}]]+\text{cyclic permutations}=0$ lead to the relation
\begin{equation}
    (m+n)(m-n)F(m,n)\delta_{m+n+l,0}=0, 
\end{equation}
which is solved by $F(m,n)=0$. Besides, it is easy to check that there is no constraint on $A(m,n)$ and that the deformation induced by $G(m,n)=constant$ is a formal deformation. Therefore, we have the new deformed algebra
\begin{equation}\label{Deform-JJ-1}
    \begin{split}
    &[\mathcal{J}_{m},\mathcal{J}_{n}]=\nu(m-n)\mathcal{P}_{m+n}+(m-n)A(m,n),\\
    &[\mathcal{J}_{m},\mathcal{P}_{n}]=m\delta_{m+n,0},\\
        &[\mathcal{P}_{m},\mathcal{P}_{n}]=0,
    \end{split}
\end{equation}
where $\nu$ is an arbitrary constant. This is the most formal deformation of \thh \ in its commutator $[\mathcal{J},\mathcal{J}]$. For $\nu\neq0$, \eqref{Deform-JJ-1} can be simplified by the redefinition $\mathcal{P}_{m}\rightarrow \mathcal{P}_{m}-\frac{A(m,n)}{\nu}$, leading to 
\begin{equation}\label{MostDeform-JJ}
    \boxed{  \begin{split}
    &[\mathcal{J}_{m},\mathcal{J}_{n}]=\nu(m-n)\mathcal{P}_{m+n} \ ,\\
    &[\mathcal{J}_{m},\mathcal{P}_{n}]=m\delta_{m+n,0} \ ,\\
        &[\mathcal{P}_{m},\mathcal{P}_{n}]=0 \ ,
    \end{split}}
\end{equation}
which we called ${\mathfrak{H}_{3}}_{\nu}$.
\paragraph{Remark.} One should note that algebra \eqref{MostDeform-JJ} can be obtained as contraction of two Virasoro algebras, as we discuss in detail in Appendix \ref{app_Heisenbergfromcontractions}. Besides, when we consider linear central term in the commutator $[\mathcal{J},\mathcal{P}]$ of the $W(0,b)$ algebra \eqref{Deform-G,F,BarG}, the algebra \eqref{MostDeform-JJ} is obtained from its contraction. 

\subsubsection*{$\bullet$ Deformations of \texorpdfstring{$[\mathcal{J},\mathcal{P}]$}{JP}}

Next, we consider deformations of the commutator $[\mathcal{J},\mathcal{P}]$ as
\begin{equation}\label{Deform-JP}
    \begin{split}
    &[\mathcal{J}_{m},\mathcal{J}_{n}]=0 \ ,\\
    &[\mathcal{J}_{m},\mathcal{P}_{n}]=m\delta_{m+n,0}+\Bar{F}(m,n)\mathcal{J}_{m+n}+\Bar{G}(m,n)\mathcal{P}_{m+n}+\Bar{A}(m,n) \ ,\\
        &[\mathcal{P}_{m},\mathcal{P}_{n}]=0 \ .
    \end{split}
\end{equation}
The Jacobi identities 
$[\mathcal{J}_{m},[\mathcal{J}_{n},\mathcal{P}_{l}]]+\text{cyclic permutations}=0$ lead to the relations
\begin{equation}\label{Bar(G,F)}
    \Bar{G}(n,l)\Bar{F}(m,n+l)-\Bar{G}(m,l)\Bar{F}(n,l+m)=0 \ ,
\end{equation}
\begin{equation}\label{Bar(G,G)}
    \Bar{G}(n,l)\Bar{G}(m,n+l)-\Bar{G}(m,l)\Bar{G}(n,l+m)=0 \ ,
\end{equation}
\begin{equation}\label{Bar(G,A)}
   (m\Bar{G}(n,l)-n\Bar{G}(m,l))\delta_{m+n+l,0}+\Bar{G}(n,l)\Bar{A}(m,n+l)-\Bar{G}(m,l)\Bar{A}(n,l+m)=0 \ .
\end{equation}
On the other hand, the Jacobi identities $[\mathcal{P}_{m},[\mathcal{P}_{n},\mathcal{J}_{l}]]+\text{cyclic permutations}=0$ lead to
\begin{equation}\label{Bar(F,F)}
    \Bar{F}(l,n)\Bar{F}(n+l,m)-\Bar{F}(l,m)\Bar{F}(l+m,n)=0 \ ,
\end{equation}
\begin{equation}\label{Bar(F,G)}
    \Bar{F}(l,n)\Bar{G}(n+l,m)-\Bar{F}(l,m)\Bar{G}(l+m,n)=0 \ ,
\end{equation}
\begin{equation}\label{Bar(F,A)}
    (\Bar{F}(l,n)(n+l)-\Bar{F}(l,m)(l+m))\delta_{m+n+l,0}+\Bar{F}(l,n)\Bar{A}(n+l,m)-\Bar{F}(l,m)\Bar{A}(l+m,n)=0\ .
\end{equation}

Focusing on infinitesimal deformation means that we should consider separately the relations with first order in functions, which are $(m\Bar{G}(n,l)-n\Bar{G}(m,l))\delta_{m+n+l,0}=0$ and $(\Bar{F}(l,n)(n+l)-\Bar{F}(l,m)(l+m))\delta_{m+n+l,0}=0$. The first relation is solved by $\Bar{G}(m,n)=\alpha m+\beta f(m) \delta_{m+n,0}$ while the second leads to $\Bar{F}(m,n)=\alpha n+\beta g(n) \delta_{m+n,0}+\gamma (m-n)$. By plugging these infinitesimal solutions into \eqref{Bar(G,F)}-\eqref{Bar(G,A)} and \eqref{Bar(F,F)}-\eqref{Bar(F,A)}, one finds that they are solved only by $\Bar{G}(m,n)=\alpha m$, $\Bar{G}(m,n)=\beta m^k \delta_{m+n,0}$, $\Bar{F}(m,n)=\gamma n$ or $\Bar{F}(m,n)=\eta n^k \delta_{m+n,0}$, where $k\in\mathbb{Z}_+$ is a positive integer. In fact, each of the solutions of $\bar{G}(m,n)$ leads to an independent formal deformation. The same takes place for solutions of $\bar{F}(m,n)$ such that we find four independent formal deformations \footnote{We also realize that relation \eqref{Bar(G,A)} leads to $\Bar{A}(m,n)=\Bar{\varepsilon} m\delta_{m+n,0}$, which just modifies the value of central term and will not be considered as a new algebra.}. Thus, we find new algebras through deformation procedure by $\bar{G}(m,n)$ 
\begin{equation}\label{Deform-JP-barG}
\boxed{
    \begin{split}
    &[\mathcal{J}_{m},\mathcal{J}_{n}]=0 \ ,\\
    &[\mathcal{J}_{m},\mathcal{P}_{n}]=m\,\delta_{m+n,0}+(\alpha\, m\,+\beta\,m^k\, \delta_{m+n,0})\mathcal{P}_{m+n} \ ,
    \\
        &[\mathcal{P}_{m},\mathcal{P}_{n}]=0 \ ,
    \end{split}}
\end{equation}
 and new algebras induced by $\bar{F}(m,n)$ 
 \begin{equation}\label{Deform-JP-barF}
 \boxed{
    \begin{split}
    &[\mathcal{J}_{m},\mathcal{J}_{n}]=0 \ ,\\
    &[\mathcal{J}_{m},\mathcal{P}_{n}]=m\delta_{m+n,0}+(\gamma\,n\,+\eta\, n^k\, \delta_{m+n,0})\mathcal{J}_{m+n} \ ,\\
        &[\mathcal{P}_{m},\mathcal{P}_{n}]=0 \ ,
    \end{split}}
\end{equation}
where $\alpha$, $\beta$, $\gamma$ and $\eta$ are four independent parameters, corresponding to four independent formal deformations which cannot be generally turned on simultaneously. For future use, we denote the algebra \eqref{Deform-JP-barG} by ${\mathfrak{H}_{3}}_{\alpha}$ when $\beta=0$.

 \paragraph{Remark.} For the deformation induced by $\beta=0$, one can easily check that, by a rescaling $\mathcal{J}_{m} \rightarrow m\, \mathcal{J}_{m}$ in \eqref{Deform-JP-barG}, a new commutator $ [\mathcal{J}_{m},\mathcal{P}_{n}]=\delta_{m+n,0}+\alpha \mathcal{P}_{m+n}$ is obtained. One can further introduce another redefinition $\mathcal{P}_{m} \to \mathcal{P}_{m}-\frac{\delta_{m,0}}{\alpha}$  \footnote{Note that for $\alpha=0$ this redefinition is singular and we really just have the Heisenberg algebra. If we naively take $\alpha=0$ in \eqref{Deform-JP-barG22}, then we clearly are not anymore in the Heisenberg branch. As a consequence, the value $\alpha=0$ cannot be taken in \eqref{Deform-JP-barG22}.} to obtain a new algebra 
 \begin{equation}\label{Deform-JP-barG22}
    \begin{split}
    &[\mathcal{J}_{m},\mathcal{J}_{n}]=0 \ ,\\
    &[\mathcal{J}_{m},\mathcal{P}_{n}]=\alpha\, \mathcal{P}_{m+n} \ ,\\
        &[\mathcal{P}_{m},\mathcal{P}_{n}]=0 \ .
    \end{split}
\end{equation}
In fact, we can rescale $\mathcal{J}_{m}$ again such that $\alpha=1$, which unveils that we are dealing with a discrete deformation. This means that all $\alpha\neq0$ are really equivalent, so the only inequivalent algebras are $\alpha=0$ (undeformed Heisenberg) and $\alpha\neq0$ \eqref{Deform-JP-barG22}. It is also worth to point out that the algebra \eqref{Deform-JP-barG} can be also obtained as a contraction of $W(0,b)$ algebra if $\beta=0$.

In section \ref{sect_weylbms}, we will discuss that \eqref{Deform-JP-barG22} is part of the larger $\mathfrak{bmsw}_3$ algebra \eqref{eq:weylbms3} and will play an important role connecting the latter to $\mathfrak{witt}\oplus\mathfrak{H}_3$ \eqref{Heisenberg-direct-sum-algebra-3d}.  


\subsubsection*{$\bullet$ Deformations of \texorpdfstring{$[\mathcal{P},\mathcal{P}]$}{JP}}
Finally, we investigate deformations of the commutator $[\mathcal{P},\mathcal{P}]$
\begin{equation}\label{Deform-PP}
    \begin{split}
    &[\mathcal{J}_{m},\mathcal{J}_{n}]=0 \ ,\\
    &[\mathcal{J}_{m},\mathcal{P}_{n}]=m\delta_{m+n,0} \ ,\\
        &[\mathcal{P}_{m},\mathcal{P}_{n}]=(m-n)\tilde{F}(m,n)\mathcal{J}_{m+n}+(m-n)\tilde{G}(m,n)\mathcal{P}_{m+n}+(m-n)\tilde{A}(m,n) \ ,
    \end{split}
\end{equation}
where $\tilde{F}(m,n)$, $\tilde{G}(m,n)$ and $\tilde{A}(m,n)$ are symmetric functions. The Jacobi identities $[\mathcal{P}_{m},[\mathcal{P}_{n},\mathcal{P}_{l}]]+\text{cyclic permutations}=0$ yield 
\begin{multline}\label{tilde(G,G)}
(n-l)(m-n-l)\tilde{G}(n,l)\tilde{G}(m,n+l)+(l-m)(n-l-m)\tilde{G}(l,m)\tilde{G}(n,l+m)+ \\
(m-n)(l-m-n)\tilde{G}(m,n)\tilde{G}(l,m+n)=0 \ ,
\end{multline}
\begin{multline}\label{tilde(G,F)}
(n-l)(m-n-l)\tilde{G}(n,l)\tilde{F}(m,n+l)+(l-m)(n-l-m)\tilde{G}(l,m)\tilde{F}(n,l+m)+ \\
(m-n)(l-m-n)\tilde{G}(m,n)\tilde{F}(l,m+n)=0 \ ,
\end{multline}
and
\begin{multline}\label{tilde(G,F,A)}
(n+l-m)(n-l)\tilde{G}(n,l)\tilde{A}(l+n,m)+(l+m-n)(l-m)\tilde{G}(l,m)\tilde{A}(l+m,n)\\+(m+n-l)(m-n)\tilde{G}(m,n)\tilde{A}(m+n,l)\\
+\{(n+l)(n-l)\tilde{F}(n,l)+(l+m)(l-m)\tilde{F}(l,m)+(m+n)(m-n)\tilde{F}(m,n)\}\delta_{m+n+l,0}=0 \ .
\end{multline}
Then we consider the Jacobi identities $[\mathcal{P}_{m},[\mathcal{P}_{n},\mathcal{J}_{l}]]+\text{cyclic permutations}=0$, which lead to
\begin{equation}
    (m+n)(m-n)\tilde{G}(m,n)=0 \ , 
\end{equation}
implying $\tilde{G}(m,n)=0$. Following the same argument mentioned before, the relation $\{(n+l)(n-l)\tilde{F}(n,l)+(l+m)(l-m)\tilde{F}(l,m)+(m+n)(m-n)\tilde{F}(m,n)\}\delta_{m+n+l,0}=0$, which is first order in functions, leads to $\tilde{F}(m,n)=constant$. Besides, it is easy to check that there is no constraint on $\tilde{A}(m,n)$, even though it can be absorbed by a redefinition (when $\eta\neq0$), and that $\tilde{F}(m,n)=constant\equiv\eta$ induces a formal deformation  
\begin{equation}\label{Deform-PP-1}
\boxed{
    \begin{split}
    &[\mathcal{J}_{m},\mathcal{J}_{n}]=0 \ ,\\
    &[\mathcal{J}_{m},\mathcal{P}_{n}]=m\delta_{m+n,0} \ ,\\
        &[\mathcal{P}_{m},\mathcal{P}_{n}]=\eta(m-n)\mathcal{J}_{m+n} \ .
    \end{split}}
\end{equation}
We note that, by exchanging $\mathcal{J}_{m}$ and $\mathcal{P}_{m}$ in \eqref{Deform-PP-1}, relation \eqref{MostDeform-JJ} will be obtained. As a consequence, both algebras are isomorphic. 


\subsection{General deformations of $\mathfrak{H}_3$}
\label{sect_Heisenbergtogether}
Now, we investigate the deformations of \eqref{Dirac-3d} in all the commutators simultaneously
\begin{equation}\label{eq:moregendefh3}
    \begin{split}
    &[\mathcal{J}_{m},\mathcal{J}_{n}]=(m-n)F(m,n)\mathcal{J}_{m+n}+(m-n)G(m,n)\mathcal{P}_{m+n}+(m-n)A(m,n) \ ,\\
    &[\mathcal{J}_{m},\mathcal{P}_{n}]=m\delta_{m+n,0}+\Bar{F}(m,n)\mathcal{J}_{m+n}+\Bar{G}(m,n)\mathcal{P}_{m+n}+\Bar{A}(m,n) \ ,\\
            &[\mathcal{P}_{m},\mathcal{P}_{n}]=(m-n)\tilde{F}(m,n)\mathcal{J}_{m+n}+(m-n)\tilde{G}(m,n)\mathcal{P}_{m+n}+(m-n)\tilde{A}(m,n) \ ,
    \end{split}
\end{equation}
where $F(m,n)$, $G(m,n)$, $A(m,n)$, $\tilde{F}(m,n)$, $\tilde{G}(m,n)$ and $\tilde{A}(m,n)$ are symmetric functions.

As the detailed analysis of the constraints coming from the Jacobi identities is cumbersome and not very instructive, we relegate it to appendix \ref{app_JacobiH3}. From this analysis, we obtain the following possibilities for the deformations:

\begin{enumerate}
    \item The deformation induced by $G(m,n)=constant$ and $\tilde{F}(m,n)=constant$, when the other deformations are turned off, does not lead to a formal deformation since the last term in both relations \eqref{JJP-Coeff-J} and \eqref{PPJ-Coeff-P} implies that $G$ or $\tilde{F}$ have to be zero. The corresponding algebras are either \eqref{Deform-JJ-1} or \eqref{Deform-PP-1}. 
    
    
    \item The deformation induced by $\Bar{G}(m,n)=\alpha m-\beta n$ and $F(m,n)=constant=\beta$, when the other deformations are turned off, leads to a formal deformation. Relations \eqref{JJJ-Coeff-J} and \eqref{JJP-Coeff-P} are satisfied with the mentioned solutions. 
    The corresponding algebra reads 
    \begin{equation}\label{Deform-F,BarG}
    \boxed{
    \begin{split}
    &[\mathcal{J}_{m},\mathcal{J}_{n}]=\beta(m-n)\mathcal{J}_{m+n} \ ,\\
    &[\mathcal{J}_{m},\mathcal{P}_{n}]=m\delta_{m+n,0}-\beta(bm+n)\mathcal{P}_{m+n} \ ,\\
        &[\mathcal{P}_{m},\mathcal{P}_{n}]=0 \ ,
    \end{split}}
\end{equation}
    where $b=-\frac{\alpha}{\beta}$. Note that we can choose the latter since $\alpha$ and $\beta$ are two arbitrary parameters.  
    The same deformation will be obtained when one considers $\tilde{G}(m,n)=constant=\alpha$ and $\Bar{F}(m,n)=\alpha m-\beta n$ and uses the fact that the algebra is symmetric under the exchange $\mathcal{J}\leftrightarrow \mathcal{P}$. 
    
    Relations \eqref{FAG-JJ} and \eqref{Bar(G-A-F)-A-G-tildeA} lead to obtain the central extension of $W(0,b)$, denoted as $\widehat{W}(0,b)$, for generic value of $b$ as
   \begin{equation}\label{central-W(0,b)}
    \boxed{
    \begin{split}
    &[\mathcal{J}_{m},\mathcal{J}_{n}]=\beta(m-n)\mathcal{J}_{m+n} +(\alpha m^3-{\alpha'} m)\delta_{m+n,0}\ ,\\
    &[\mathcal{J}_{m},\mathcal{P}_{n}]=m\delta_{m+n,0}-\beta(bm+n)\mathcal{P}_{m+n} \ ,\\
        &[\mathcal{P}_{m},\mathcal{P}_{n}]=0 \ . 
    \end{split} }
\end{equation}
    
     On the other hand, for specific values $b=\{-1,0,1 \}$, one can find new central terms in other commutators. The new central terms as solutions of \eqref{FAG-JJ}, \eqref{Bar(G-A-F)-A-G-tildeA} and \eqref{Bar(F,A,G)-tilde(A,F,G)} give rise, respectively, to the algebras 
    \begin{equation}\label{eq:W-1centralext}
    \begin{split}
    &[\mathcal{J}_{m},\mathcal{J}_{n}]=\beta(m-n)\mathcal{J}_{m+n}+(\alpha m^3-{\alpha'} m)\delta_{m+n,0} \ ,\\
    &[\mathcal{J}_{m},\mathcal{P}_{n}]=m\delta_{m+n,0}+\beta(m-n)\mathcal{P}_{m+n}+(\Bar{\alpha}m^3-\bar{\alpha}'m)\delta_{m+n,0} \ ,\\
    &[\mathcal{P}_{m},\mathcal{P}_{n}]=0 \ ,
    \end{split}
\end{equation}
    \begin{equation}\label{eq:W00centralext}
    \begin{split}
    &[\mathcal{J}_{m},\mathcal{J}_{n}]=\beta(m-n)\mathcal{J}_{m+n}+(\alpha m^3-{\alpha'} m)\delta_{m+n,0} \ ,\\
    &[\mathcal{J}_{m},\mathcal{P}_{n}]=m\delta_{m+n,0}+\beta(-n)\mathcal{P}_{m+n}+(\Bar{\alpha}m^2+\bar{\alpha}'m)\delta_{m+n,0} \ ,\\
    &[\mathcal{P}_{m},\mathcal{P}_{n}]=\tilde{\alpha}m\delta_{m+n,0} \ ,
    \end{split}
\end{equation}
and 
\begin{equation}\label{hatW01}
    \begin{split}
    &[\mathcal{J}_{m},\mathcal{J}_{n}]=\beta(m-n)\mathcal{J}_{m+n}+(\alpha m^3-{\alpha'} m)\delta_{m+n,0} \ ,\\
    &[\mathcal{J}_{m},\mathcal{P}_{n}]=m\delta_{m+n,0}-\beta(m+n)\mathcal{P}_{m+n}+(\Bar{\alpha}m+\bar{\alpha}')\delta_{m+n,0} \ ,\\
    &[\mathcal{P}_{m},\mathcal{P}_{n}]=0 \ ,
    \end{split}
\end{equation}

which is in agreement with the theorem 1.2 of \cite{gao2011low}. 

\paragraph{Remark.} When $b\neq 1$, we note that, by means of the redefinition $P_{m}:= \mathcal{P}_{m}-\frac{\delta_{m,0}}{\beta(b-1)}$, the linear central term can be absorbed in \eqref{Deform-F,BarG}. A further rescaling of the generators $\mathcal{J}_m\to\beta\mathcal{J}_m$ and $P_{m}\to\frac{P_{m}}{\beta}$ shows that indeed \eqref{Deform-F,BarG} is isomorphic to $W(0;b)$ for $b\neq1$. Furthermore, rescaling $\mathcal{J}_m\to\alpha\mathcal{J}_m$, it is clear that \eqref{eq:W-1centralext}, \eqref{eq:W00centralext} and \eqref{hatW01} correspond respectively to $\widehat{W}(0;-1)$, $\widehat{W}(0;0)$ and $\widehat{W}(0;1)$.

It is noteworthy to point out that, although we start with the Heisenberg non-trivial central extension, new algebras are obtained after deformation in which central terms that are trivial and can be absorbed by redefinition pop up. This shows that the deformation procedure can change the role of a non-trivial central term to a trivial one.

  \item The deformation induced by $G(m,n)=constant=\nu$ and $\Bar{G}(m,n)=\alpha m-\beta n$ and $F(m,n)=constant=\beta$, when other deformations are turned off, leads to a formal deformation. Relations \eqref{JJJ-Coeff-J}, \eqref{JJJ-Coeff-P} and \eqref{JJP-Coeff-P} are satisfied with the mentioned solutions. The corresponding algebra reads 
    \begin{equation}\label{Deform-G,F,BarG}
    \boxed{\begin{split}
    &[\mathcal{J}_{m},\mathcal{J}_{n}]=\beta(m-n)\mathcal{J}_{m+n}+\nu(m-n)\mathcal{P}_{m+n} \ ,\\
    &[\mathcal{J}_{m},\mathcal{P}_{n}]=m\delta_{m+n,0}-\beta(bm+n)\mathcal{P}_{m+n} \ ,\\
        &[\mathcal{P}_{m},\mathcal{P}_{n}]=0 \ ,
    \end{split}}
\end{equation}
where again we have $b=-\frac{\alpha}{\beta}$. The new algebra is denoted by $\widehat{W}_{\nu}(0,b)$ and it is one of the two deformation mother algebras of $\mathfrak{H}_3$. This is a three-parametric family of algebras which can be centrally extended to become four-parametric with a central charge in the $[\mathcal{J},\mathcal{J}]$ commutator. The algebras \eqref{MostDeform-JJ}, \eqref{Deform-JP-barG} and \eqref{Deform-G,BarG} can be obtained from various choices of the parameters in $\widehat{W}_{\nu}(0,b)$

The same deformation will be obtained when one considers $\tilde{G}(m,n)=constant=\beta$ and $\Bar{F}(m,n)=\alpha m-\beta n$ and $\tilde{F}(m,n)=constant=\nu$ and uses the fact that the algebra is symmetric under the exchange $\mathcal{J}\leftrightarrow \mathcal{P}$.
In the algebra \eqref{Deform-G,F,BarG} one can make usage of the redefinition $\mathcal{P}_{m}\rightarrow P_{m}+\frac{\delta_{m,0}}{\beta(b-1)}$ to obtain
\begin{equation}
    \begin{split}
    &[\mathcal{J}_{m},\mathcal{J}_{n}]=\beta(m-n)\mathcal{J}_{m+n}+\nu(m-n){P}_{m+n}+\nu(m-n)(\frac{1}{\beta(b-1)}\delta_{m+n,0}) \ ,\\
    &[\mathcal{J}_{m},{P}_{n}]=-\beta(bm+n)P_{m+n} \ ,\\
        &[{P}_{m},{P}_{n}]=0 \ .
    \end{split}
\end{equation}

    Another redefinition $\mathcal{J}_{m}\rightarrow J_{m}+\frac{1}{\beta}(\frac{-\nu}{\beta(b-1)}\delta_{m,0})$, followed by the rescaling $J_{m}\rightarrow \beta J_{m}$, leads to
      \begin{equation}\label{Deform-G,F,BarG-1}
    \begin{split}
    &[{J}_{m},{J}_{n}]=(m-n){J}_{m+n}+\frac{\nu}{\beta^2}(m-n){P}_{m+n} \ ,\\
    &[{J}_{m},{P}_{n}]=-(bm+n)P_{m+n} \ ,\\
        &[{P}_{m},{P}_{n}]=0 \ ,
    \end{split}
\end{equation}
    where the procedure is only valid for $b\neq 1$.
     Next, we can make use of the redefinition ${J}_{m}\rightarrow\Bar{J}_{m}-\frac{\nu}{\beta^2 b}P_{m} $ to obtain 
     \begin{equation}\label{Deform-G,F,BarG-2}
    \begin{split}
    &[\Bar{J}_{m},\Bar{J}_{n}]=(m-n)\Bar{J}_{m+n} \ ,\\
    &[\Bar{J}_{m},{P}_{n}]=-(bm+n)P_{m+n} \ ,\\
        &[{P}_{m},{P}_{n}]=0 \ ,
    \end{split}
\end{equation}
    which is only valid for $b\neq0$. 
    On the other hand, one finds that the $\nu$-term in \eqref{Deform-G,F,BarG}, when $b=1$, can be absorbed by a redefinition  
       $ \mathcal{J}_{m}\rightarrow J_{m}-\frac{\nu}{\beta}P_{m}-\frac{\nu}{\beta^2}\delta_{m,0}$ and $\mathcal{P}_{m}\rightarrow P_{m}$. The resultant algebra is given by 
    \begin{equation}\label{Deform-G,F,BarG-3}
    \begin{split}
    &[{J}_{m},{J}_{n}]=(m-n){J}_{m+n}\ ,\\
    &[{J}_{m},{P}_{n}]=-(m+n)P_{m+n}+m\delta_{m+n,0} \ ,\\
        &[{P}_{m},{P}_{n}]=0 \ .
    \end{split}
\end{equation}
    In this way, we have found three independent algebras: \eqref{Deform-G,F,BarG-3} when $b=1$, \eqref{Deform-G,F,BarG-1} when $b=0$ and \eqref{Deform-G,F,BarG-2} for other values of $b$. Note that the algebra \eqref{Deform-G,F,BarG-2} corresponds to $W(0;b\neq\{0,1\})$, while the algebras \eqref{Deform-G,F,BarG-1} (for $b=0$) and \eqref{Deform-G,F,BarG-3} (for $b=1$) can be understood, respectively, as a deformation of $W(0;0)$ and a central extension of $W(0;1)$. 
    

The central terms as solutions of \eqref{Bar(G-A-F)-A-G-tildeA}, \eqref{FAG-JJ} and \eqref{Bar(F,A,G)-tilde(A,F,G)} for \eqref{Deform-G,F,BarG} when $b=\{-1,1 \}$ give us, respectively, the algebras
    \begin{equation}\label{central-Deform-G,BarG}
    \begin{split}
    &[\mathcal{J}_{m},\mathcal{J}_{n}]=\alpha(m-n)\mathcal{J}_{m+n}+\nu(m-n)\mathcal{P}_{m+n}+(\eta\, m^3-\beta\, m)\delta_{m+n,0} \ ,\\
    &[\mathcal{J}_{m},\mathcal{P}_{n}]=m\delta_{m+n,0}+\alpha(m-n)\mathcal{P}_{m+n}+(\Bar{\eta}\,m^3-\Bar{\beta}\,m)\delta_{m+n,0} \ ,\\
        &[\mathcal{P}_{m},\mathcal{P}_{n}]=0 \ ,
    \end{split}
\end{equation}
and
    \begin{equation}\label{central-Deform-G,BarG-1}
    \begin{split}
  &[\mathcal{J}_{m},\mathcal{J}_{n}]=\alpha(m-n)\mathcal{J}_{m+n}+\nu(m-n)\mathcal{P}_{m+n}+(\eta\, m^3-\beta\, m)\delta_{m+n,0} \ ,\\
    &[\mathcal{J}_{m},\mathcal{P}_{n}]=m\delta_{m+n,0}-\alpha(m+n)\mathcal{P}_{m+n}+(\Bar{\eta}\,m+\Bar{\beta})\delta_{m+n,0} \ ,\\
        &[\mathcal{P}_{m},\mathcal{P}_{n}]=0 \ .
    \end{split}
\end{equation}
From the last term of \eqref{Bar(G-A-F)-A-G-tildeA}, we observe that the $\nu$-term cannot appear when $b=0$, since it admits a central term in its last commutator \footnote{One may consider the case $\tilde{\alpha}=0$ in \eqref{eq:W00centralext} to obtain a $\nu$-term as a non-trivial deformation of $\widehat{W}(0,0)$, which cannot be absorbed by a change of the basis.}. 


One can easily notice that, in the algebra \eqref{central-Deform-G,BarG}, the linear central terms can be absorbed by a redefinition of generators in a similar line as described before. Besides, the $\nu$-term can be absorbed by a redefinition as $\mathcal{J}_{m}\rightarrow J_{m}+\frac{\nu}{\alpha}{P}_{m}$ and $\mathcal{P}_{m}\rightarrow P_{m}$. This does not change $\Bar{\eta}$ while shifts ${\eta}$ to $\Tilde{\eta}={\eta}-\frac{2\nu}{\alpha}\Bar{\eta}$. The new algebra has the form
 \begin{equation}\label{central-Deform-G,BarG-3}
    \begin{split}
    &[{J}_{m},{J}_{n}]=\alpha(m-n){J}_{m+n}+(\tilde{\eta}\, m^3)\delta_{m+n,0} \ ,\\
    &[{J}_{m},{P}_{n}]=\alpha(m-n){P}_{m+n}+(\Bar{\eta}\,m^3)\delta_{m+n,0} \ ,\\
        &[{P}_{m},{P}_{n}]=0 \ .
    \end{split}
\end{equation}
In this way, the $\nu$-term, which changes the value of central term, can be interpreted as a quantum correction in asymptotically flat spacetime analysis \cite{Merbis:2019wgk}. 

It turns out that the $\nu$-term and the linear central term in the first commutator of \eqref{central-Deform-G,BarG-1} can also be reabsorbed by the redefinition $\mathcal{J}_{m}\rightarrow J_{m}-\frac{\nu}{\alpha}P_{m}-2\frac{\nu}{\alpha^2}(1+\Bar{\eta})\delta_{m,0}+\frac{\beta}{2\alpha}\delta_{m+n,0}$.





\paragraph{Remark 1.} Relation \eqref{linear-BarG-F} is also solved by $F(m,n)=0$ and $\Bar{G}(m,n)=\alpha m$. The only relation that should be checked is \eqref{JJJ-Coeff-P}, which is satisfied by $\Bar{G}(m,n)=\alpha m$ and $G(m,n)=constant=\nu$, leading to the new algebra
 \begin{equation}\label{Deform-G,BarG}\boxed{
    \begin{split}
    &[\mathcal{J}_{m},\mathcal{J}_{n}]=\nu(m-n)\mathcal{P}_{m+n} \ ,\\
    &[\mathcal{J}_{m},\mathcal{P}_{n}]=m\delta_{m+n,0}+\alpha m \mathcal{P}_{m+n} \ ,\\
        &[\mathcal{P}_{m},\mathcal{P}_{n}]=0 \ ,
    \end{split}}
\end{equation}

which we denote by ${\mathfrak{H}_{3}}_{\nu\alpha}$. 

For $\alpha\neq0$, and using the redefinition as $\mathcal{J}\to\mathcal{J}+\frac{\nu}{\alpha} \mathcal{P}$ and $\mathcal{P}\rightarrow \mathcal{P}$, we obtain 
\begin{equation}\label{Deform-G,BarG-new}
    \begin{split}
    &[\mathcal{J}_{m},\mathcal{J}_{n}]=\beta\,m\delta_{m+n} \ ,\\
    &[\mathcal{J}_{m},\mathcal{P}_{n}]=m\delta_{m+n,0}+\alpha m \mathcal{P}_{m+n} \ ,\\
        &[\mathcal{P}_{m},\mathcal{P}_{n}]=0 \ ,
    \end{split}
    \end{equation}
    where $\beta=-\frac{2\nu}{\alpha}$. It is obvious that algebra \eqref{Deform-G,BarG-new} can be obtained as a specific contraction of $\widehat{W}(0,b)$. 
A further redefinition $\mathcal{P}_m\rightarrow \mathcal{P}_m-\frac{\delta_{m,0}}{\alpha}$ leads to 
    \begin{equation}\label{jjdelta}
    \begin{split}
    &[\mathcal{J}_{m},\mathcal{J}_{n}]={\beta m \delta_{m+n.0}}\ ,\\
    &[\mathcal{J}_{m},\mathcal{P}_{n}]=\alpha m \mathcal{P}_{m+n} \ ,\\
        &[\mathcal{P}_{m},\mathcal{P}_{n}]=0 \ .
    \end{split}
\end{equation}
This algebra can be understood as a central extension of \eqref{Deform-JP-barG22} in the $[\mathcal{J},\mathcal{J}]$ commutator.

\paragraph{Remark 2.} Another formal deformation is generated by $F(m,n)=0$ and $\bar{G}(m,n)=\alpha \,m^k\,\delta_{m+n,0}$ for the values of $k=1,3$. In the case $k=1$, the algebra reads as follows after suitable redefinition
\begin{equation}
    \begin{split}
    &[\mathcal{J}_{m},\mathcal{J}_{n}]=\frac{\nu}{1+\alpha}(m-n)\mathcal{P}_{m+n} \ ,\\
    &[\mathcal{J}_{m},\mathcal{P}_{n}]=m\, \mathcal{P}_0\, \delta_{m+n,0} \ ,\\
        &[\mathcal{P}_{m},\mathcal{P}_{n}]=0 \ .
\    \end{split}
\end{equation}
This algebra can be viewed as a combination of \eqref{MostDeform-JJ} and \eqref{Deform-JP-barG} when $\alpha=0$ and $k=1$.

 \item The deformation induced by $G(m,n)=constant$ and $\Bar{F}(m,n)=\alpha(m-n)$ and $\tilde{G}(m,n)=constant$, when other deformations are turned off, leads to a formal deformation. Relations \eqref{JJJ-Coeff-J}, \eqref{JJP-Coeff-P} and \eqref{PPJ-coeff-J} are satisfied with the mentioned solutions. The other solutions of $\Bar{F}(m,n)$, like $\Bar{F}(m,n)=-\alpha n$ and $\Bar{F}(m,n)=-\alpha (m+n)$, cannot satisfy relation \eqref{JJP-Coeff-P} so they do not lead to any new algebra, in agreement with our previous result. The corresponding algebra reads 
    \begin{equation}\label{Deform-G,BarF,TildeG}\boxed{
    \begin{split}
    &[\mathcal{J}_{m},\mathcal{J}_{n}]=\nu(m-n)\mathcal{P}_{m+n} \ ,\\
    &[\mathcal{J}_{m},\mathcal{P}_{n}]=m\delta_{m+n,0}+\alpha(m-n)\mathcal{J}_{m+n} \ ,\\
        &[\mathcal{P}_{m},\mathcal{P}_{n}]=\alpha(m-n)\mathcal{P}_{m+n} \ ,
    \end{split}}
\end{equation} 
which, when the linear central term can be absorbed by redefinition of generator $\mathcal{J}$, consists of just two copies of the Witt algebra. 
The same deformation can be obtained when considering $\tilde{F}(m,n)=constant=\nu$, $\Bar{G}(m,n)=\alpha(m-n)$ and $F(m,n)=constant=\alpha$, using the fact that the algebra is symmetric under the exchange $\mathcal{J}\leftrightarrow\mathcal{P}$. 

\paragraph{Remark.}\label{Poincare}
It is worth to point out that the linear central term in \eqref{Deform-G,BarF,TildeG} is a trivial central term, in the sense that it can be absorbed by an appropriate redefinition of generators, while the central term in $\mathfrak{H}_{3}$ we started with is a non-trivial central term. Indeed, deformation and contraction procedures can change the role of generators. This is analogous to the deformation/contraction relation between Poincaré and Bargmann algebras. In fact, to obtain Bargmann algebra through contraction of Poincaré algebra a new $\mathfrak{u}(1)$ generator, as trivial central term, has to be added to Poincaré algebra \cite{Andringa:2010it}. This trivial central term after contraction becomes a non-trivial central term in the Bargmann algebra. The same takes place for the $\nu$-term in \eqref{MostDeform-JJ} and, for instance, \eqref{Deform-G,F,BarG}. In fact, in \eqref{MostDeform-JJ} the $\nu$-term is a non-trivial part of the algebra but after deformation for $b\neq\{0 ,1\}$ it can be absorbed by a redefinition.

    The central terms as solutions of \eqref{Bar(G-A-F)-A-G-tildeA}, \eqref{FAG-JJ}, \eqref{Bar(F,A,G)-tilde(A,F,G)} and \eqref{tilde(F,G,A)-BarA} give rise to the algebra
     \begin{equation}\label{Deform-central-G,BarF,TildeG}\boxed{
    \begin{split}
    &[\mathcal{J}_{m},\mathcal{J}_{n}]=\nu(m-n)\mathcal{P}_{m+n}+(\alpha' m^3-\beta m)\delta_{m+n,0} \ ,\\
    &[\mathcal{J}_{m},\mathcal{P}_{n}]=m\delta_{m+n,0}+\alpha(m-n)\mathcal{J}_{m+n}+(\Bar{\alpha} m^3-\bar{\beta} m)\delta_{m+n,0} \ ,\\
        &[\mathcal{P}_{m},\mathcal{P}_{n}]=\alpha(m-n)\mathcal{P}_{m+n}+(\tilde{\alpha} m^3-\tilde{\beta} m)\delta_{m+n,0} \ ,
    \end{split}}
\end{equation} 

with the constraints $-\nu\tilde{\beta}+\alpha\beta=0$ and $-\nu\tilde{\alpha}+\alpha\alpha'=0$. These constraints, together with the possibility of absorbing two linear central terms via redefinition of $\mathcal{J}_m$ and $\mathcal{P}_m$, translate into the known fact that there are just two independent central terms for two copies of Virasoro $\mathfrak{vir}\oplus\mathfrak{vir}$.  


     \item The deformation induced by $F(m,n)=constant=\alpha$, $\Bar{G}(m,n)=\alpha(m-n)$, $\tilde{G}(m,n)=constant=\beta$ and $\Bar{F}(m,n)=\beta(m-n)$, when other deformations are turned off, does not lead to a formal deformation since relations \eqref{PPJ-Coeff-P} and \eqref{JJP-Coeff-J} are not satisfied with the mentioned solutions turned on simultaneously. 
     
     
    \item The deformation induced by $F(m,n)=constant=\alpha$, $G(m,n)=constant$, $\Bar{F}(m,n)=\eta(m-n)$, $\Bar{G}(m,n)=\alpha(m-n)$, $\tilde{F}(m,n)=constant$ and $\tilde{G}(m,n)=constant=\eta$, when they are turned on simultaneously, yields a formal deformation, since all relations obtained through Jacobi identities are satisfied. 
The corresponding algebra reads
    \begin{equation}\label{Deform-G,F,BarF,BarG,TildeF,TildeG}
    \boxed{
    \begin{split}
    &[\mathcal{J}_{m},\mathcal{J}_{n}]=\alpha(m-n)\mathcal{J}_{m+n}+\nu(m-n)\mathcal{P}_{m+n} \ ,\\
    &[\mathcal{J}_{m},\mathcal{P}_{n}]=m\delta_{m+n,0}+\eta(m-n)\mathcal{J}_{m+n}+\alpha(m-n)\mathcal{P}_{m+n} \ ,\\
        &[\mathcal{P}_{m},\mathcal{P}_{n}]=\zeta(m-n)\mathcal{J}_{m+n}+\eta(m-n)\mathcal{P}_{m+n} \ , 
    \end{split}}
\end{equation} 
with the constraint $\alpha\eta-\nu\zeta=0$. This algebra turns out to be a three-parametric deformation mother algebra which we will denote by $\mathcal{H}_3(\alpha,\eta,\nu)$ and contains $\mathfrak{bms}_3$, $\mathfrak{witt}\oplus\mathfrak{witt}$ and $\mathfrak{witt}\oplus\mathfrak{u}(1)$ for several choices of the parameters.


For example, let us explore the simple case $\alpha=\eta=\nu=\zeta$ where we denote them as $\varepsilon$. Making use of the redefinitions $\mathcal{J}_{m}\rightarrow {P}_{m}-{J}_{m}$ and $\mathcal{P}_{m}\rightarrow {P}_{m}+{J}_{m}$, we obtain
 \begin{equation}
 \boxed{
    \begin{split}
    &[{J}_{m},{J}_{n}]=-\frac{m}{2}\delta_{m+n,0} \ ,\\
    &[{P}_{m},{J}_{n}]=0 \ ,\\
        &[{P}_{m},{P}_{n}]=2\varepsilon(m-n){P}_{m+n}+\frac{m}{2}\delta_{m+n,0} \ .
    \end{split}}
\end{equation} 

This corresponds to a direct sum of the Witt algebra with a current algebra, which can be also obtained from contraction of two Virasoro algebras. 



    
     The central terms as solutions of \eqref{Bar(G-A-F)-A-G-tildeA}, \eqref{FAG-JJ}, \eqref{Bar(F,A,G)-tilde(A,F,G)} and \eqref{tilde(F,G,A)-BarA} give rise to the algebra
     \begin{equation}\label{Deform-G,F,BarF,BarG,TildeF,TildeG-centralext}
     \boxed{
    \begin{split}
    &[\mathcal{J}_{m},\mathcal{J}_{n}]=\alpha(m-n)\mathcal{J}_{m+n}+\nu(m-n)\mathcal{P}_{m+n}+(\alpha'' m^3-\beta'' m)\delta_{m+n,0} \ ,\\
    &[\mathcal{J}_{m},\mathcal{P}_{n}]=m\delta_{m+n,0}+\eta(m-n)\mathcal{J}_{m+n}+\alpha(m-n)\mathcal{P}_{m+n}+(\Bar{\alpha} m^3-\bar{\beta} m)\delta_{m+n,0} \ ,\\
        &[\mathcal{P}_{m},\mathcal{P}_{n}]=\zeta(m-n)\mathcal{J}_{m+n}+\eta(m-n)\mathcal{P}_{m+n}+(\tilde{\alpha} m^3-\tilde{\beta} m)\delta_{m+n,0} \ ,
    \end{split}}
\end{equation} 
with the constraints $\alpha\eta-\nu\zeta=0$, $-\eta\beta''+\nu\tilde{\beta}=0$, $-\eta\alpha''+\nu\tilde{\alpha}=0$, $-\zeta\beta''+\alpha\tilde{\beta}=0$ and $-\zeta\alpha''+\alpha\tilde{\alpha}=0$. This algebra turns out to be a five-parametric deformation mother algebra which we will denote by $\widehat{\mathcal{H}_3}(\alpha,\eta,\nu)$ and which contains $\widehat{\mathfrak{bms}_3}$, $\mathfrak{vir}\oplus\mathfrak{vir}$ and $\mathfrak{vir}\oplus\mathfrak{u}(1)$ for several choices of the parameters. In fact, it corresponds to the centrally extended $\mathcal{H}_3(\alpha,\eta,\nu)$ mother algebra \eqref{Deform-G,F,BarF,BarG,TildeF,TildeG}.


For example, let us investigate the simple case $\alpha=\nu=\eta=\zeta=\alpha''=\beta''=\bar{\alpha}=\bar{\beta}=\tilde{\alpha}=\tilde{\beta}=\sigma$. By means of the redefinitions $\mathcal{J}_{m}\rightarrow {P}_{m}-{J}_{m}$ and $\mathcal{P}_{m}\rightarrow {P}_{m}+{J}_{m}$, we learn that the RHS is equivalent to that of \eqref{Deform-G,F,BarF,BarG,TildeF,TildeG} together with the addition of a central term $\sigma( m^3- m)\delta_{m+n,0}$
\begin{equation}\label{Virasoropluscurrent}
\boxed{
    \begin{split}
    &[{J}_{m},{J}_{n}]=-\frac{m}{2}\delta_{m+n,0} \ ,\\
    &[{P}_{m},{J}_{n}]=0 \ ,\\
        &[{P}_{m},{P}_{n}]=2\sigma(m-n){P}_{m+n}+\frac{m}{2}\delta_{m+n,0}+\sigma( m^3- m)\delta_{m+n,0} \ .
    \end{split}}
\end{equation} 

The linear central terms in the last commutator can also be absorbed and one concludes that \eqref{Virasoropluscurrent} is just a direct sum of Virasoro and a current algebra denoted as $\mathfrak{vir}\oplus\mathfrak{u}(1)$.


\end{enumerate}

\subsubsection{Summary of new algebras and disjoint families}
\label{sect_summaryalgebras}

\begin{figure}[h!]
    \centering
\begin{tikzpicture}[node distance=1cm, auto]
 \node[punkt] (H3) {\hyperref[Dirac-3d]{$\mathfrak{H}_{3}$}};
  \node[right=5.2cm of H3] (dummy) {};
  
  \node[punkt,right=5.2cm of dummy] (twovir) {\hyperref[Deform-central-G,BarF,TildeG]{$\mathfrak{vir}\oplus\mathfrak{vir}$}};
  
 \node[punkt, inner sep=6pt,above=4cm of dummy] (bms3) {\hyperref[eq:W-1centralext]{$\widehat{\mathfrak{bms}}_{3}$}};

 \node[punkt, inner sep=6pt,above=2.5cm of dummy] (nu) {\hyperref[MostDeform-JJ]{${\mathfrak{H}_{3}}_{\nu}$}};

 
 \node[punkt,above=1cm of dummy] (W) {\hyperref[Deform-G,F,BarG]{$\widehat{W}_{\nu}(0,b)$}};

 \node[punkt, inner sep=6pt,below=2.5cm of dummy](a) {\hyperref[Deform-JP-barG]{${\mathfrak{H}_{3}}_{\alpha}$} };
 


 \node[punkt,below=1cm of dummy] (anu) {\hyperref[Deform-G,BarG]{${\mathfrak{H}_{3}}_{\nu\alpha}$} };
 
 \node[punkt, inner sep=6pt,below=4cm of dummy]
 (viru) {\hyperref[Virasoropluscurrent]{$\mathfrak{vir}\oplus\mathfrak{u}(1)$}};

   
   \path [-stealth, thick]
    (H3.east) edge [bend left] node [left=0.1cm] {} (W.west);
   \path [-stealth, thick]
    (H3.east) edge [bend left] node [left=0.1cm] {} (bms3.west);
     \path [-stealth, thick]
    (H3.east) edge [bend left] node [left=0.1cm] {} (nu.west);

     \path [-stealth, thick]
    (H3.east) edge node [left=0.1cm] {} (twovir.west);
    
     \path [-stealth, thick]
    (bms3.east) edge [pil, bend left] node [left=0.1cm] {} (twovir.west);
    
     \path [-stealth, thick]
    (bms3.east) edge [pil, bend left] node [left=0.1cm] {} (W.east);
    
    \path [-stealth, thick]
    (a.east) edge [pil, bend right] node [left=0.1cm] {} (anu.east);

    \path [-stealth, thick]
    (a.east) edge [pil, bend right] node [left=0.1cm] {} (anu.east);

    \path [-stealth, thick]
    (anu.east) edge [pil, bend right] node [left=0.1cm] {} (W.east);
    
     \path [-stealth, thick]
    (a.east) edge [pil, bend right] node [left=0.1cm] {} (W.east);
    
     \path [-stealth, thick]
    (nu.east) edge [pil, bend left] node [left=0.1cm] {} (W.east);
    
    \path [-stealth, thick]
    (nu.west) edge [pil, bend left] node [left=0.1cm] {} (bms3.west);
    
     \path [-stealth, thick]
    (nu.east) edge [pil, bend left] node [left=0.1cm] {} (twovir.west);
    
     \path [-stealth, thick]
    (H3.east) edge [bend right] node [left=0.1cm] {} (a.west);
    \path [-stealth, thick]
    (H3.east) edge [bend right] node [left=0.1cm] {} (anu.west);
    \path [-stealth, thick]
    (H3.east) edge [bend right] node [left=0.1cm] {} (viru.west);

    
    \path [-stealth, thick]
    (viru.east) edge [pil, bend right] node [left=0.1cm] {} (twovir.west);
    
    \path [-stealth, thick]
    (nu.west) edge [pil, bend right] node [left=0.1cm] {} (anu.west);
   

\end{tikzpicture}
\caption{Various algebras obtained as formal deformations of $\mathfrak{H}_3$ and their connections. Each line indicates a deformation/contraction relationship. The arrows signal the direction of deformation, while the contractions follow the inverse direction.}
\label{Fig.1}
\end{figure}
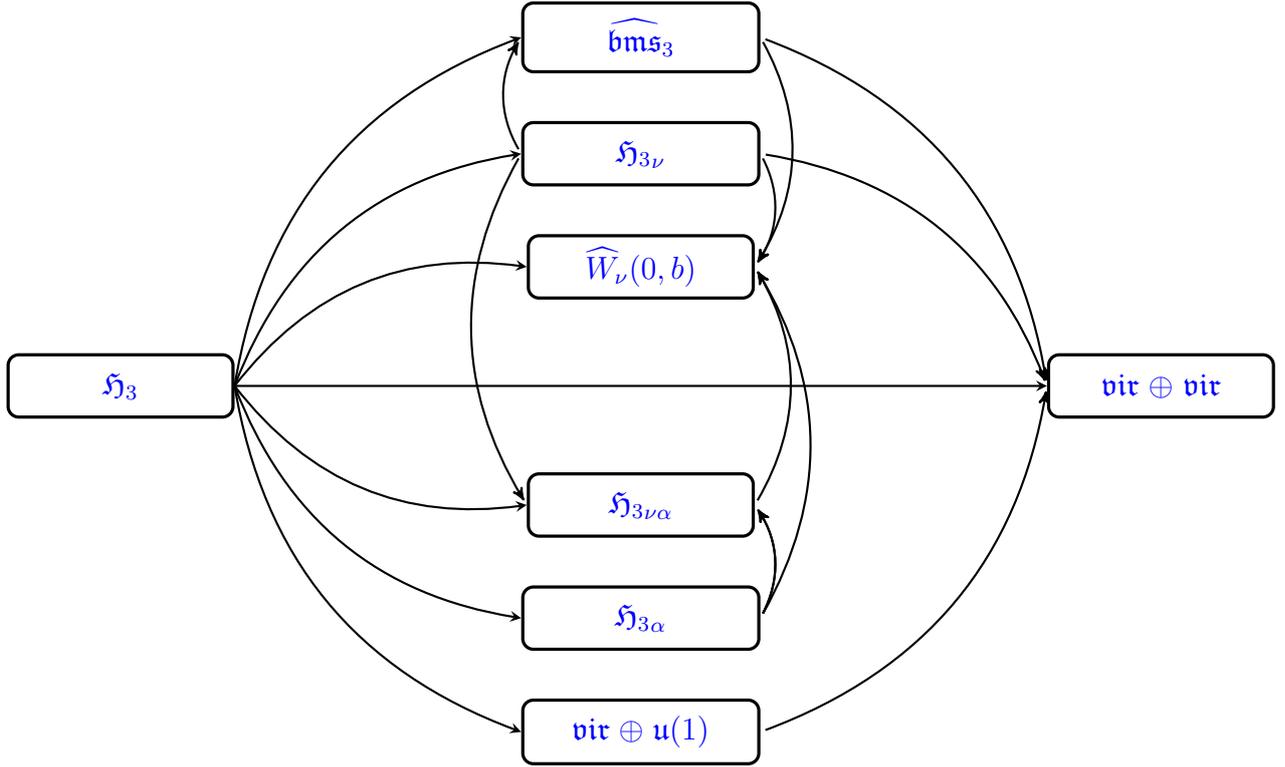

We have shown that $\mathfrak{H}_{3}$ can be deformed into various algebras. Some of these algebras are connected to each other through deformation/contraction procedures. The algebras which cannot be related in this way are called ``{\it disjoint}'' algebras. 
In figure \ref{Fig.1}, we summarize the main algebras obtained through deformations of $\mathfrak{H}_{3}$ and specify their connections \footnote{It is important to note that there are more new algebras obtained through deformation of $\mathfrak{H}_3$, as described in sections \ref{sect_H3separate} and \ref{sect_Heisenbergtogether}. Their deformation/contraction connections are worth to study in future works.}. Each line in the diagram indicates a deformation/contraction relation between two algebras, where the arrows indicate the direction of deformation. For instance, the algebra \eqref{Deform-G,BarG} cannot be deformed into neither $\mathfrak{bms}_{3}$ nor two Virasoro algebras. On the other hand, it can be deformed into \eqref{central-W(0,b)} and contracted to \eqref{MostDeform-JJ} and \eqref{Deform-JP-barG} (when $\beta=0$). 

It has been shown in \cite{FarahmandParsa:2018ojt} that two copies of the Virasoro algebra $\mathfrak{vir}\oplus\mathfrak{vir}$ are a rigid algebra, in the sense that it cannot be deformed into any non-trivial algebra. According to a notion of rigidity for family algebras introduced in \cite{Safari:2020pje}, which states that the $\widehat{W}(0,b)$ family algebra can be deformed into the $\widehat{W}(0,\bar{b})$ algebra with shifted parameters \footnote{In \cite{Safari:2020pje}, we showed that $\widehat{W}(0,b)$ cannot be deformed into $\widehat{W}(a,b)$ because of the presence of a linear central term.}, one can check that the two algebras $\widehat{W}_{\nu}(0,b)$ and $\mathfrak{vir}\oplus\mathfrak{vir}$ are disjoint algebras for generic values of $b$. It is also obvious that $\mathfrak{vir}\oplus\mathfrak{u}(1)$ can just be deformed into $\mathfrak{vir}\oplus\mathfrak{vir}$ and contracted to $\mathfrak{H}_{3}$. So it is disjoint from the other algebras in middle column of diagram. Furthermore, $\mathfrak{bms}_{3}$ and ${\mathfrak{H}_{3}}_{\nu\alpha}$ cannot be connected through deformation/contraction relation, being disjoint algebras.

As a final point, it is worth pointing out that, among the various algebras in figure \ref{Fig.1}, $\mathfrak{H}_{3\nu}$, $\mathfrak{H}_{3\alpha}$, $\mathfrak{H}_{3\nu\alpha}$ and $\mathfrak{vir}\oplus\mathfrak{u}(1)$ are algebras which have not yet been obtained as asymptotic/near horizon symmetry algebras.

\paragraph{Remark 1.} We should mention a couple of points about the diagram. Firstly, $\widehat{\mathfrak{bms}}_{3}$ can be deformed into $\widehat{W}_{\nu}(0,b)$ when $c_{JP}=0$, even though we kept the trivial linear central term in \eqref{eq:W-1centralext}. Secondly, it was discussed in the previous section that the $\nu$-deformation in $\widehat{W}_{\nu}(0,b)$ is only not trivial for $b=0$ and for other values of $b$ can be absorbed by redefinition of generators. 

\paragraph{Remark 2.} From the diagram \ref{Fig.1} we observe that all the deformations end in one of the two mother algebras $\widehat{W}_{\nu}(0,b)$ \eqref{Deform-G,F,BarG} and $\widehat{\mathcal{H}_3}(\alpha,\nu,\eta)$ \eqref{Deform-G,F,BarF,BarG,TildeF,TildeG-centralext} obtained in section \ref{sect_Heisenbergtogether}. Interestingly, the algebra $\mathfrak{vir}\oplus\mathfrak{vir}$ is a rigid representative of $\widehat{\mathcal{H}_3}(\alpha,\nu,\eta)$. In addition, $\mathfrak{bms}_3$ is a representative of both $\widehat{\mathcal{H}_3}(\alpha,\nu,\eta)$ and $\widehat{W}_{\nu}(0,b)$.

\subsubsection{Deformations vs Sugawara construction}
\label{sec_sugawara}



An alternative to deformations which relates $\mathfrak{H}_3$ with other algebras is the celebrated Sugawara construction \cite{Sugawara:1967rw,DiFrancesco:1997nk}. It has been shown in \cite{Afshar:2016kjj} that $\mathfrak{vir}$, $\mathfrak{bms}_3$ and Virasoro-Kac-Moody algebras can be obtained from Heisenberg algebras through different twisted and untwisted Sugawara constructions. On the other hand, algebras like $W(0,b)$ \eqref{eq:Wab} for some values of $b$ cannot be found in this way. In fact, the Witt algebra can be obtained through a quadratic Sugawara-like term as 
\begin{equation}
    \mathcal{L}_{m}:=\sum_{k}\mathcal{J}_{m-k}\mathcal{P}_{k},
\end{equation}
where $\mathcal{L}'s$ satisfy Witt algebra. Taking into account that the ideal part of $W(0,b)$ is abelian, its generators can be expressed in terms of generators of $\mathfrak{H}_3$ as 
\begin{equation}
    \mathcal{M}_{m}:={a}\sum_{k}\mathcal{J}_{m-k}\mathcal{J}_{k}+{b}\mathcal{J}_{m}+{c}\mathcal{P}_{m}. 
\end{equation}
One then finds that $\mathcal{L}'s$ and $\mathcal{M}'s$ satisfy $W(0,-1)$ when ${b}={c}=0$, ${a}=1$, and $W(0,0)$ when ${b}={c}=1$, ${a}=0$. But it is not possible to obtain any other solution for $b\neq\{0,-1\}$. 

A natural question to ask is whether we can obtain all the new algebras coming from $\mathfrak{H}_3$ deformations by means of a Sugawara construction. It turns out that some of the new algebras can be found in this way. For example, \eqref{MostDeform-JJ} can be obtained through the following Sugawara construction
\begin{equation}
    J_m=\frac{\nu}{2}\sum_k \mathcal{P}_{m-k}\mathcal{P}_k+a m \mathcal{P}_m + \mathcal{J}_m \ , \ \ \ P_m=\mathcal{P}_m \ .
\end{equation}

Nevertheless, it does not seem to be always possible to obtain our new algebras using such constructions. For example, let us analyze a possible Sugawara construction connecting Heisenberg to \eqref{Deform-JP-barG}. One can readily note that for a maximally quadratic ansatz of the form:
\begin{align}
    J_m=\sum_k (a_1 \mathcal{J}_{m-k}\mathcal{J}_k + b_1 \mathcal{J}_{m-k}\mathcal{P}_k + c_1 \mathcal{P}_{m-k}\mathcal{P}_k) + \text{non-quadratic} \ , \\
    P_m=\sum_k (a_2 \mathcal{J}_{m-k}\mathcal{J}_k + b_2 \mathcal{J}_{m-k}\mathcal{P}_k + c_2 \mathcal{P}_{m-k}\mathcal{P}_k) + \text{non-quadratic} \ ,
\end{align}
the commutators $[J,J]=[P,P]=0$ impose the conditions $a_i b_i=0$, $c_i b_i=0$ and $4a_ic_i+b_i^2=0$ for both $i=1,2$. As a consequence, the most general solution we can have is $b_i=0$ and $a_i\neq0$ or $c_i\neq0$. None of the four possibilities with both $J_m$ and $P_m$ having a quadratic term allows for the quadratic piece of a term $\alpha m P_m$ in the commutator $[J_m,P_n]$. Another two possibilities consist of setting any quadratic term in $P_m$ to zero and allow for quadratic term in $J_m$ as $J_m\propto \sum_k a_1 \mathcal{J}_{m-k}\mathcal{J}_k $ or $J_m\propto \sum_k c_1 \mathcal{P}_{m-k}\mathcal{P}_k $. It is easy to check that such terms cannot generate the desired $\alpha m P_m$ independently of the form of the non-quadratic terms. Finally, it is straightforward to realize that when the quadratic terms in both generators are fixed to vanish, one cannot generate $\alpha m P_m$ in the commutator $[J_m,P_n]$. As a consequence, a conventional quadratic Sugawara construction cannot connect $\mathfrak{H}_3$ with \eqref{Deform-JP-barG}.

Replicating the argument for the absence of Sugawara construction for $W(0,b\neq\{0,-1\})$, we easily realize that another example which cannot be obtained through a possible Sugawara construction is \eqref{Deform-F,BarG}.

To sum up, the deformations belonging to the family $\mathcal{H}_3(\alpha,\nu,\eta)$ \eqref{Deform-G,F,BarF,BarG,TildeF,TildeG-centralext} which crucially involve $(m-n)$ terms can be Sugawara constructed, while those coming from the family $\widehat{W}_{\nu}(0,b)$ \eqref{Deform-G,F,BarG} cannot be obtained via a conventional quadratic Sugawara construction of $\mathfrak{H}_3$ for generic values of $b$.

\section{Deformations of  \texorpdfstring{$\mathfrak{witt}\oplus \mathfrak{H}_{3}$}{Witt+Heisenberg}  }\label{sect_weylbms}

In the following, we will study the deformations of $\mathfrak{witt}\oplus \mathfrak{H}_{3}$ as described in appendix \ref{app_deformations} and, specifically, its connection with the $\mathfrak{bmsw}_3$ algebra \eqref{eq:weylbms3}, recently discussed in various works \cite{Adami:2020ugu,Donnay:2020fof,Batlle:2020hia}. 

\subsection{Deformations of \texorpdfstring{$\mathfrak{witt}\oplus \mathfrak{H}_{3}$} \ \  without central extensions}

In this subsection, we would like to study the deformations of $\mathfrak{witt}\oplus \mathfrak{H}_{3}$, excluding central terms. To this end, we deform the commutators of $\mathfrak{witt}\oplus \mathfrak{H}_{3}$ as follows \footnote{Here we take back the notation used in \eqref{eq:moregendefh3} for the $\mathfrak{H}_{3}$ part without the central terms.}: 
\begin{equation}\label{eq:witth3gendefwithoutcentral}
\begin{split}
    &[\mathcal{L}_{m},\mathcal{L}_{n}]=(m-n)\mathcal{L}_{m+n}+A_1(m,n)\mathcal{L}_{m+n}+B_1(m,n)\mathcal{J}_{m+n}+C_1(m,n)\mathcal{P}_{m+n} \ ,\\
    &[\mathcal{L}_{m},\mathcal{J}_{n}]=A_2(m,n)\mathcal{L}_{m+n}+B_2(m,n)\mathcal{J}_{m+n}+C_2(m,n)\mathcal{P}_{m+n} \ ,\\
    &[\mathcal{L}_{m},\mathcal{P}_{n}]=A_3(m,n)\mathcal{L}_{m+n}+B_3(m,n)\mathcal{J}_{m+n}+C_3(m,n)\mathcal{P}_{m+n} \ ,\\
    &[\mathcal{P}_{m},\mathcal{P}_{n}]=A_4(m,n)\mathcal{L}_{m+n}+(m-n)\tilde{F}(m,n)\mathcal{J}_{m+n}+(m-n)\tilde{G}(m,n)\mathcal{P}_{m+n} \ , \\
    &[\mathcal{J}_{m},\mathcal{J}_{n}]=A_5(m,n)\mathcal{L}_{m+n}+(m-n)F(m,n)\mathcal{J}_{m+n}+(m-n)G(m,n)\mathcal{P}_{m+n} \ ,\\
    &[\mathcal{J}_{m},\mathcal{P}_{n}]=m\delta_{m+n,0}+A_6(m,n)\mathcal{L}_{m+n}+\Bar{F}(m,n)\mathcal{J}_{m+n}+\Bar{G}(m,n)\mathcal{P}_{m+n} \ ,
\end{split}
\end{equation}
where $A_{i}$, $B_{i}$ and $C_{i}$ are arbitrary functions to be determined by the Jacobi identities. First of all, we consider infinitesimal deformations such that we should only keep linear order in the functions. Besides, we can use the fact that $\mathfrak{witt}$ is a rigid subalgebra \cite{fialowski2012formal}, which allows us to set $A_{1}(m,n)=0$ in \eqref{eq:witth3gendefwithoutcentral}. The Jacobi identities $[\mathcal{L}_{m},[\mathcal{L}_{n},\mathcal{J}_{l}]]+\text{cyclic permutations}=0$ yield
\begin{multline}
    (m-n) B_2(m+n,l) \mathcal{J}_{l + m + n} +(m-n) C_2(m+n,l) \mathcal{P}_{l + m + n}-lC_{1}(m,n)\delta_{l+m+n,0} \\ +[-(l + m - n) A_2(m, l) + (l - m + n) A_2(n, l) + (m - n) A_2(m + n, l) ] \mathcal{L}_{l + m + n} = 0 \ ,
\end{multline}
which leads to obtain $B_2(m,n)=C_2(m,n)=C_1(m,n)=0$ and a constraint for $A_{2}(m,n)=\alpha (m-n)$.

The Jacobi identities $[\mathcal{L}_{m},[\mathcal{L}_{n},\mathcal{P}_{l}]]+\text{cyclic permutations}=0$ lead to 
\begin{multline}
    (m-n) B_3(m + n, l) \mathcal{J}_{l + m + n} +(m-n) C_3(m + n, l) \mathcal{P}_{l + m + n} -lB_{1}(m,n)\delta_{l+m+n,0} \\ +[-(l + m - n) A_3(m, l) + (l - m + n) A_3(n, l) + (m - n) A_3(m + n, l) ] \mathcal{L}_{l + m + n}= 0 \ ,
\end{multline}
which forces $B_3(m,n)=C_3(m,n)=B_1(m,n)=0$ and a constraint for $A_{3}(m,n)=\beta (m-n)$.

Similarly, the Jacobi identities $[\mathcal{L}_{m},[\mathcal{P}_{n},\mathcal{P}_{l}]]+\text{cyclic permutations}=0$ and $[\mathcal{L}_{m},[\mathcal{J}_{n},\mathcal{J}_{l}]]+\text{cyclic permutations}=0$ cause $A_{4}(m,n)=A_{5}(m,n)=0$. Lastly, the Jacobi identities $[\mathcal{L}_{m},[\mathcal{J}_{n},\mathcal{P}_{l}]]+\text{cyclic permutations}=0$ lead to $A_{6}(m,n)=0$. It follows, as a consequence, that the Jacobi identities for the $\mathfrak{H}_{3}$ sector decouple from $\mathcal{L}$ and are exactly the same as in section \ref{sec_defsHeisenbeg}.

Therefore, the unique new infinitesimal deformations with respect to section \ref{sec_defsHeisenbeg} are those given by $A_2(m,n)=\alpha (m-n)$ and $A_3(m,n)=\beta (m-n)$. These new infinitesimal deformations can only be formal if they also satisfy the non-linear equations coming from the Jacobi identities $[\mathcal{L}_{m},[\mathcal{J}_{n},\mathcal{J}_{l}]]+\text{cyclic permutations}=0$ , $[\mathcal{L}_{m},[\mathcal{P}_{n},\mathcal{P}_{l}]]+\text{cyclic permutations}=0$ and $[\mathcal{L}_{m},[\mathcal{J}_{n},\mathcal{P}_{l}]]+\text{cyclic permutations}=0$. It turns out that they are only involved in the following 
\begin{equation}\label{eq:newnonlinearjacobis}
\begin{split}
    & (l-n)(l-m+n)[\alpha^2-\alpha F(n,l)-\beta G(n,l)] \mathcal{L}_{l + m + n} = 0 \  , \\
    & (l-n)(l-m+n)[\beta^2-\beta \tilde{G}(n,l)-\alpha \tilde{F}(n,l)] \mathcal{L}_{l + m + n} = 0 \ , \\
    & (l-m+n)[\alpha\beta (l-n)+\beta \bar{G}(n,l)+\alpha \bar{F}(n,l)] \mathcal{L}_{l + m + n} = 0 \ . 
    \end{split}
\end{equation}
Thus, in order to find new formal deformations from $A_2(m,n)=\alpha (m-n)$ and/or $A_3(m,n)=\beta (m-n)$, we will have to check if any of the formal deformations found in section \ref{sec_defsHeisenbeg} satisfies also \eqref{eq:newnonlinearjacobis} for $\alpha\neq0$ and/or $\beta\neq0$. Let us analyze the different possibilities:

\begin{enumerate}
    \item { $\alpha\neq 0$ while $\beta=0$.} To obtain a non-trivial solution, the relation \eqref{eq:newnonlinearjacobis} implies that $\bar{F}(m,n)=\tilde{F}(m,n)=0$ while $F(m,n)=\alpha$. So all the algebras obtained through deformations of $\mathfrak{H}_{3}$ in previous section which satisfy these constraints can be considered as deformations of $\mathfrak{witt} \oplus \mathfrak{H}_{3}$ induced by the function $A_{2}(m,n)=\alpha (m-n)$. For instance, we can obtain this new algebra
      \begin{equation}
    \begin{split}
     &[\mathcal{L}_{m},\mathcal{L}_{n}]=(m-n)\mathcal{L}_{m+n} \ ,\\
    &[\mathcal{L}_{m},\mathcal{J}_{n}]=\alpha\,(m-n)\mathcal{L}_{m+n} \ ,\\
    &[\mathcal{L}_{m},\mathcal{P}_{n}]=0 \ ,\\
    &[\mathcal{J}_{m},\mathcal{J}_{n}]=\alpha\,(m-n)\mathcal{J}_{m+n} \ ,\\
    &[\mathcal{J}_{m},\mathcal{P}_{n}]=m\delta_{m+n,0}-\alpha\,(bm+n)\mathcal{P}_{m+n} \ ,\\
        &[\mathcal{P}_{m},\mathcal{P}_{n}]=0.
    \end{split}
\end{equation}
One can check that, after redefinition of the generators, this algebra is nothing but $\mathfrak{witt}\oplus W(0;b)$.
    
    \item { $\beta\neq 0$ while $\alpha=0$.} To obtain a non-trivial solution, the relation \eqref{eq:newnonlinearjacobis} implies that $\bar{G}(m,n)=\tilde{G}(m,n)=0$ while $G(m,n)=\beta$. These constraints lead to obtain the same algebras as in the previous case by using the replacement $\mathcal{J}\leftrightarrow \mathcal{P}$.

   \item { $\alpha= \beta\neq 0$.} There are various options 
  \begin{equation}
    \begin{split}
     &[\mathcal{L}_{m},\mathcal{L}_{n}]=(m-n)\mathcal{L}_{m+n} \ ,\\
    &[\mathcal{L}_{m},\mathcal{J}_{n}]=\alpha\,(m-n)\mathcal{L}_{m+n} \ ,\\
    &[\mathcal{L}_{m},\mathcal{P}_{n}]=\alpha\,(m-n)\mathcal{L}_{m+n} \ ,\\
    &[\mathcal{J}_{m},\mathcal{J}_{n}]=\frac{\alpha}{2}\,(m-n)\mathcal{J}_{m+n}+\frac{\alpha}{2}\,(m-n)\mathcal{P}_{m+n} \ ,\\
    &[\mathcal{J}_{m},\mathcal{P}_{n}]=m\delta_{m+n,0}+\frac{\alpha}{2}\,(m-n)\mathcal{J}_{m+n}+\frac{\alpha}{2}\,(m-n)\mathcal{P}_{m+n} \ ,\\
        &[\mathcal{P}_{m},\mathcal{P}_{n}]=\frac{\alpha}{2}\,(m-n)\mathcal{J}_{m+n}+\frac{\alpha}{2}\,(m-n)\mathcal{P}_{m+n}.
    \end{split}
\end{equation}
We note that, after redefinition of the generators, this algebra corresponds to the direct sum of two Witt algebras and a Kac-Moody current algebra $\mathfrak{witt}\oplus\mathfrak{witt}\oplus\mathfrak{u}(1)$.

\item { $\alpha\neq 0 \neq \beta$.}  In order to satisfy \eqref{eq:newnonlinearjacobis}, the unique reasonable ansatz is given by $\tilde{F}(m,n)=G(m,n)=0$, $\tilde{G}(m,n)=\beta$, $F(m,n)=\alpha$. Three  possibilities follow: 
$\{ \bar{F}(m,n) , \bar{G}(m,n) \}=\frac{(m-n)}{2}\{ \beta , \alpha \}$ , $\{ \bar{F}(m,n) , \bar{G}(m,n) \}=(m-n)\{ \beta , 0  \}$ and 
$\{ \bar{F}(m,n) , \bar{G}(m,n) \}=(m-n)\{ 0 , \alpha  \}$. Nevertheless, the $\mathfrak{H}_3$ Jacobi identities are not satisfied unless $\alpha=0$ and/or $\beta=0$ for the three options.

\end{enumerate}

\paragraph{Remark.} We observe that all the new algebras obtained via deformation of $\mathfrak{witt}\oplus\mathfrak{H}_3$, without involving central extensions, come from deformations on the $\mathfrak{H}_3$ part. The $\mathfrak{witt}$ part of the algebra is well-known to be rigid \cite{fialowski2012formal} and all the studied mixed deformations can be redefined as deformations only in the $\mathfrak{H}_3$ part. In the next subsection, we will explore whether this pattern still holds if we allow for central extensions.


\subsection{Deformations of \texorpdfstring{$\mathfrak{witt}\oplus \mathfrak{H}_{3}$} \ \ with central extensions}

In the following, we examine the effect of adding central terms. Henceforth, we study the deformations of $\mathfrak{witt}\oplus \mathfrak{H}_{3}$, including the addition of central terms. To this end, we deform the commutators of $\mathfrak{witt}\oplus \mathfrak{H}_{3}$ as follows: 
\begin{equation}\label{eq:virh3gendefwithcentral}
\begin{split}
     &[\mathcal{L}_{m},\mathcal{L}_{n}]=(m-n)\mathcal{L}_{m+n}+A_1(m,n)\mathcal{L}_{m+n}+B_1(m,n)\mathcal{J}_{m+n}+C_1(m,n)\mathcal{P}_{m+n}+D_{1}(m,n) \ ,\\
    &[\mathcal{L}_{m},\mathcal{J}_{n}]=A_2(m,n)\mathcal{L}_{m+n}+B_2(m,n)\mathcal{J}_{m+n}+C_2(m,n)\mathcal{P}_{m+n}+D_{2}(m,n) \ ,\\
    &[\mathcal{L}_{m},\mathcal{P}_{n}]=A_3(m,n)\mathcal{L}_{m+n}+B_3(m,n)\mathcal{J}_{m+n}+C_3(m,n)\mathcal{P}_{m+n}+D_{3}(m,n) \ ,\\
    &[\mathcal{P}_{m},\mathcal{P}_{n}]=A_4(m,n)\mathcal{L}_{m+n}+(m-n)\tilde{F}(m,n)\mathcal{J}_{m+n}+(m-n)\tilde{G}(m,n)\mathcal{P}_{m+n}+(m-n)\tilde{A}(m,n) \ , \\
    &[\mathcal{J}_{m},\mathcal{J}_{n}]=A_5(m,n)\mathcal{L}_{m+n}+(m-n)F(m,n)\mathcal{J}_{m+n}+(m-n)G(m,n)\mathcal{P}_{m+n}+(m-n)A(m,n) \ ,\\
    &[\mathcal{J}_{m},\mathcal{P}_{n}]=m\delta_{m+n,0}+A_6(m,n)\mathcal{L}_{m+n}+\Bar{F}(m,n)\mathcal{J}_{m+n}+\Bar{G}(m,n)\mathcal{P}_{m+n}+\Bar{A}(m,n) \ ,
\end{split}
\end{equation}
where $A_{i}$, $B_{i}$, $C_{i}$ and $D_{i}$ ($i=1-6$) are arbitrary functions, whose form is constrained by the Jacobi identities. Such an analysis is straightforward, although lengthy and tedious. As a consequence, we relegate the details to appendix \ref{app_JacobiWittH3}.

We do not aim for an exhaustive analysis of all possible deformations in this case. Nevertheless, we would like to discuss several interesting different possibilities for formal deformations:
\begin{enumerate}
     \item When all the deformations in the commutators $[\mathcal{L},\mathcal{L}]$, $[\mathcal{L},\mathcal{J}]$ and $[\mathcal{L},\mathcal{P}]$ are turned off except for $D_1(m,n)=\sigma(m^3-m)\delta_{m+n,0}$ (as well as $A_4(m,n)=A_5(m,n)=A_6(m,n)=0$), we obtain formal deformations. This means that all the deformations of $\mathfrak{H}_3$ studied in section \ref{sec_defsHeisenbeg} are compatible deformations of $\mathfrak{witt}\oplus\mathfrak{H}_{3}$, even when the $\mathfrak{witt}$ subalgebra is also centrally deformed. In this way, we can deform $\mathfrak{witt}\oplus\mathfrak{H}_{3}$ to, among others, the direct sum of three Virasoro algebras or to the direct sum of centrally extended $W(0,b)$ and the Virasoro algebra. The latter algebra, for $b=-1$, has been realized in \cite{Adami:2020xkm} as the asymptotic symmetry algebra of Maxwell Chern-Simons gravity theory in 3D by considering certain boundary conditions, whereas the former has been realized in \cite{Concha:2018jjj} as an asymptotic symmetry algebra of AdS-Lorentz Chern-Simons gravity. 

    \item For the case $\alpha\neq0$ while $\beta=0$. A possibility that solves Jacobi identities is:
    \begin{equation}
\begin{split}\label{C_1,D_1,D_2,F,barG,barA}
     &[\mathcal{L}_{m},\mathcal{L}_{n}]=(m-n)\mathcal{L}_{m+n}+\alpha (m-n)\mathcal{P}_{m+n}+(c_1 m^3 - c_2 m)\delta_{m+n,0} \ ,\\
    &[\mathcal{L}_{m},\mathcal{J}_{n}]=\alpha (m-n)\mathcal{L}_{m+n}+\alpha n \delta_{m+n,0} \ ,\\
    &[\mathcal{L}_{m},\mathcal{P}_{n}]=0 \ ,\\
    &[\mathcal{P}_{m},\mathcal{P}_{n}]=0 \ , \\
    &[\mathcal{J}_{m},\mathcal{J}_{n}]=\alpha(m-n)\mathcal{J}_{m+n} \ ,\\ 
    &[\mathcal{J}_{m},\mathcal{P}_{n}]=m\delta_{m+n,0}+\alpha (m-n)\mathcal{P}_{m+n}+(c_1 m^3 - c_2 m)\delta_{m+n,0} \ .
\end{split}
\end{equation}
One can check that, after redefinition of the generators, this algebra is nothing but $\mathfrak{vir} \oplus \widehat{\mathfrak{bms}_3}$.
Similar algebras can be obtained for the case $\beta\neq0$ and $\alpha=0$, by using the replacement $\mathcal{P}\leftrightarrow \mathcal{J}$.

   \item For the case $\alpha=\beta\neq0$. The following deformation solves the Jacobi identities:
   \begin{equation}
    \begin{split}
     &[\mathcal{L}_{m},\mathcal{L}_{n}]=(m-n)\mathcal{L}_{m+n}+\gamma(m-n)\mathcal{J}_{m+n}+\gamma(m-n)\mathcal{P}_{m+n} \ ,\\
    &[\mathcal{L}_{m},\mathcal{J}_{n}]=\alpha\,(m-n)\mathcal{L}_{m+n}+\gamma n\delta_{m+n,0} \ ,\\
    &[\mathcal{L}_{m},\mathcal{P}_{n}]=\alpha\,(m-n)\mathcal{L}_{m+n}+\gamma n\delta_{m+n,0} \ ,\\
    &[\mathcal{J}_{m},\mathcal{J}_{n}]=\frac{\alpha}{2}\,(m-n)\mathcal{J}_{m+n}+\frac{\alpha}{2}\,(m-n)\mathcal{P}_{m+n} \ ,\\
    &[\mathcal{J}_{m},\mathcal{P}_{n}]=m\delta_{m+n,0}+\frac{\alpha}{2}\,(m-n)\mathcal{J}_{m+n}+\frac{\alpha}{2}\,(m-n)\mathcal{P}_{m+n} \ ,\\
        &[\mathcal{P}_{m},\mathcal{P}_{n}]=\frac{\alpha}{2}\,(m-n)\mathcal{J}_{m+n}+\frac{\alpha}{2}\,(m-n)\mathcal{P}_{m+n} \ .
    \end{split}
\end{equation}

We note that, after redefinition of the generators, this algebra corresponds to the direct sum of two Witt algebras with a Kac-Moody current algebra $\mathfrak{witt} \oplus \mathfrak{witt} \oplus \mathfrak{u}(1)$ .

\end{enumerate}
Although there might be more possibilities leading to new non-trivial algebras, we do not aim to classify all of them here. 

\paragraph{Remark 1.} It is noteworthy to point out that, through deformation of $\mathfrak{witt}\oplus\mathfrak{H}_{3}$, we can obtain the direct sum of two Virasoro algebras with a Kac-Moody current algebra $\mathfrak{vir}\oplus\mathfrak{vir}\oplus\mathfrak{u}(1)$. This algebra has been realized as asymptotic symmetry algebra of AdS$_3$ when extra Weyl symmetry is considered \cite{Alessio:2020ioh,Geiller:2021vpg}. In fact, as it is discussed in \cite{Adami:2020ugu}, when the Weyl symmetry is taken into account, various algebras, which are related to each other through slicing procedure, can be obtained. Also, one can obtain $\mathfrak{vir}\oplus\mathfrak{vir}\oplus\mathfrak{u}(1)$ by the Sugawara construction of $\mathfrak{vir}\oplus\mathfrak{H}_{3}$. 

\paragraph{Remark 2.} We observe again that all the new algebras obtained via deformation of $\mathfrak{witt}\oplus\mathfrak{H}_3$, allowing for central extensions, come from deformations on the $\mathfrak{H}_3$ part. The $\mathfrak{witt}$ part of the algebra is well-known to be rigid \cite{fialowski2012formal} and all the studied mixed deformations can be redefined as deformations only in the $\mathfrak{H}_3$ part. This example serves us to speculate that, for infinite-dimensional algebras $\mathfrak{g_1}\oplus\mathfrak{g_2}$ in which $\mathfrak{g}_1$ is rigid, all the deformations come from the $\mathfrak{g_2}$ part and, therefore, the mixed commutators remain untouched \footnote{We are not aware if this statement holds for finite-dimensional algebras and it would be definitely interesting to explore it.}. Examples in favor of this statement are the deformations obtained in this work for $\mathfrak{witt}\oplus\mathfrak{H}_3$ and the fact that $\mathfrak{vir}\oplus\mathfrak{vir}$ is rigid \cite{FarahmandParsa:2018ojt}.

\subsection{Relation between $\mathfrak{witt}\oplus\mathfrak{H}_3$ and $\mathfrak{bmsw}_3$}

The results within this section rule out the possibility of relating $\mathfrak{witt}\oplus \mathfrak{H}_{3}$ and $\mathfrak{bmsw}_3$ through a direct continuous deformation. Another way to visualize this result is to recall the fact that there is no infinitesimal linear deformation connecting $\mathfrak{witt}$ plus Abelian algebra with the $\mathfrak{bms}_3$ algebra. The same argumentation discards the possibility of relating $\mathfrak{witt}\oplus \mathfrak{H}_{3}$ and the algebra (3.34) in \cite{Adami:2020ugu} through a direct continuous deformation. 

Nevertheless, let us point out that we can relate $\mathfrak{witt}\oplus \mathfrak{H}_{3}$ and $\mathfrak{bmsw}_3$ by means of a contraction followed by a double deformation. The first step is it to contract $\mathfrak{witt}$ to an Abelian subalgebra, which can be easily achieved by a redefinition $\mathcal{L}\to \varepsilon\mathcal{L}$ and taking the limit $\varepsilon\to\infty$. Next, we deform and redefine the $\mathfrak{H}_{3}$ subalgebra to \eqref{Deform-JP-barG22}. Finally, the new formal deformation
\begin{equation}\label{eq:toWeyl-BMS}
    \begin{split}
    &[\mathcal{L}_{m},\mathcal{L}_{n}]=\beta(m-n)\mathcal{L}_{m+n} \ ,\\
    &[\mathcal{L}_{m},\mathcal{P}_{n}]=\beta (m-n) \mathcal{P}_{m+n} \ ,\\
    &[\mathcal{L}_{m},\mathcal{J}_{n}]=-\beta n \mathcal{J}_{m+n}\ ,\\ 
    &[\mathcal{J}_{m},\mathcal{J}_{n}]=0 \ ,\\
    &[\mathcal{J}_{m},\mathcal{P}_{n}]=\alpha \mathcal{P}_{m+n} \ ,  \\
        &[\mathcal{P}_{m},\mathcal{P}_{n}]=0 
    \end{split}
\end{equation}
contacts $\mathfrak{bmsw}_3$ when $\alpha$, $\beta$ are reabsorbed by redefinition of the generators. Naturally, the inverse procedure from $\mathfrak{bmsw}_3$ to $\mathfrak{witt}\oplus \mathfrak{H}_{3}$ requires the opposite steps, meaning a contraction from \eqref{eq:toWeyl-BMS} to \eqref{Deform-JP-barG22}, then a redefinition followed by a contraction to obtain $\mathfrak{H}_{3}$ plus Abelian ideal and, finally, deforming the latter to $\mathfrak{witt}\oplus \mathfrak{H}_{3}$.

An alternative way to find the $\mathfrak{bmsw}_3$ algebra is to use a different basis as a starting point. One may start with the Heisenberg-like basis (twisted double Heisenberg algebra) as \footnote{We are not aware of any discussions of this algebra in the asymptotic/boundary symmetries literature and it is intriguing to investigate its potential role in that context.} 
\begin{equation}\label{eq:toWeyl-BMS1}
    \begin{split}
    &[\mathcal{L}_{m},\mathcal{P}_{n}]=c \,m\,\delta_{m+n,0} \ ,\\
    &[\mathcal{J}_{m},\mathcal{P}_{n}]=\Bar{c} \delta_{m+n,0}\ ,  
    \end{split}
\end{equation}
where the other commutators vanish. The algebra \eqref{eq:toWeyl-BMS1} can be deformed into 
\begin{equation}
    \begin{split}
    &[\mathcal{L}_{m},\mathcal{L}_{n}]=\alpha (m-n)\mathcal{L}_{m+n} \ ,\\
    &[\mathcal{L}_{m},\mathcal{P}_{n}]=\alpha(\beta\,m-n)\mathcal{P}_{m+n}+c\,m\,\delta_{m+n,0} \ ,\\
    &[\mathcal{L}_{m},\mathcal{J}_{n}]=\alpha(-n)\mathcal{J}_{m+n}\ ,\\ 
    &[\mathcal{J}_{m},\mathcal{J}_{n}]=0 \ ,\\
    &[\mathcal{J}_{m},\mathcal{P}_{n}]= \alpha\,\mathcal{P}_{m+n}+\Bar{c}\,\delta_{m+n,0}\ ,  \\
        &[\mathcal{P}_{m},\mathcal{P}_{n}]=0 \ ,
    \end{split}
\end{equation}
where $\alpha$ is deformation parameter and the Jacobi identities force $\beta=\frac{c}{\bar{c}}-1$. This algebra is equivalent to (3.34) of \cite{Adami:2020ugu} for $s\neq0$. It should be remarked that, for $\beta \neq -1$, the central terms can be absorbed by a redefinition of generators. For specific value $\beta=1$, which implies that $\Bar{c}=\frac{c}{2}$, we find $\mathfrak{bmsw}_3$ as
\begin{equation}\label{eq:toWeyl-BMS2}
    \begin{split}
    &[\mathcal{L}_{m},\mathcal{L}_{n}]=\alpha (m-n)\mathcal{L}_{m+n} \ ,\\
    &[\mathcal{L}_{m},\mathcal{P}_{n}]=\alpha(m-n)\mathcal{P}_{m+n}+c\,m\,\delta_{m+n,0} \ ,\\
    &[\mathcal{L}_{m},\mathcal{J}_{n}]=\alpha(-n)\mathcal{J}_{m+n}\ ,\\ 
    &[\mathcal{J}_{m},\mathcal{J}_{n}]=0 \ ,\\
    &[\mathcal{J}_{m},\mathcal{P}_{n}]= \alpha\,\mathcal{P}_{m+n}+\frac{c}{2}\,\delta_{m+n,0}\ ,  \\
        &[\mathcal{P}_{m},\mathcal{P}_{n}]=0 \ .
    \end{split}
\end{equation}
 One can show that the central terms can be absorbed by redefinition as $\mathcal{P}_{m}\rightarrow P_{m}-\frac{c}{2\alpha}\delta_{m,0}$. Here we are tackling with a similar situation as we discussed in the remark below equation \eqref{Deform-G,BarF,TildeG}. Non-trivial central terms become trivial after deformation. 

 We also note that the algebra \eqref{eq:toWeyl-BMS1} can be deformed into
\begin{equation}
    \begin{split}
    &[\mathcal{L}_{m},\mathcal{L}_{n}]=\alpha (m-n)\mathcal{L}_{m+n} \ ,\\
    &[\mathcal{L}_{m},\mathcal{P}_{n}]=\alpha(-m-n)\mathcal{P}_{m+n}+c\,m\,\delta_{m+n,0} \ ,\\
    &[\mathcal{L}_{m},\mathcal{J}_{n}]=\alpha(-n)\mathcal{J}_{m+n}\ ,\\ 
    &[\mathcal{J}_{m},\mathcal{J}_{n}]=0 \ ,\\
    &[\mathcal{J}_{m},\mathcal{P}_{n}]= \Bar{c}\,\delta_{m+n,0}\ ,  \\
        &[\mathcal{P}_{m},\mathcal{P}_{n}]=0 \ ,
    \end{split}
\end{equation}
which is exactly the same as the algebra (3.34) introduced in \cite{Adami:2020ugu} when $s=0$.  
Another deformation of \eqref{eq:toWeyl-BMS1} (when $c=0$) leads to get $\mathfrak{witt}\oplus \mathfrak{H}_{3}$ as 
\begin{equation}
    \begin{split}
    &[\mathcal{L}_{m},\mathcal{L}_{n}]=\alpha (m-n)\mathcal{L}_{m+n} \ ,\\
     &[\mathcal{J}_{m},\mathcal{P}_{n}]= \Bar{c}\,\delta_{m+n,0}\ , 
    \end{split}
\end{equation}
where the other commutators are zero. 




\section{Summary and concluding remarks}\label{sec-discussions}

In this paper, we delved into boundary Heisenberg algebras and their corresponding deformations. These algebras which we briefly listed in section \ref{sect_review} and described in appendix \ref{app_spacetimeGNS} appear predominantly as boundary symmetry algebras in diverse solution phase space slicings and play a major role in the description of the spacetime structure near generic null surfaces \cite{Afshar:2016kjj, Afshar:2016wfy,Grumiller:2019fmp,Adami:2020ugu,Adami:2021sko,Adami:2021kvx,Adami:2021nnf}. As discussed in the introduction, the analysis of deformations of such algebras is expected to provide a better comprehension of the freedom in the choices of boundary conditions, when dealing with boundary analysis of the same loci, and also of the interpolation between algebras coming from the symmetry analysis at different loci. Resolving this issue might turn out to be a fundamental step for a better understanding of the black hole information puzzle. In fact, recent studies \cite{Hawking:2016msc,Averin:2016ybl,Lust:2017gez,Grumiller:2020vvv} suggest that soft/surface charges associated to diffeomorphisms in gravity may, at least partially, codify the microstrate structure of these mysterious entities. Our main objective along this work is to obtain the most general deformations of boundary Heisenberg-like algebras and relate them to other asymptotic and boundary symmetry algebras, establishing purely algebraic structures and relationships. In the long term, we expect the latter to unveil a deeper mathematical structure, guiding us in the selection of boundary conditions and degrees of freedom. 

In section \ref{sec_defsHeisenbeg}, we investigated the deformations of the infinite dimensional Heisenberg algebra $\mathfrak{H}_3$ \eqref{Dirac-3d}. Firstly, we explored deformations of isolated commutators and then we obtained the deformations of the entire algebra. We have also summarized some of the most relevant obtained new algebras and their relationship through deformation/contraction procedure in figure \ref{Fig.1}. Moreover, we briefly compared this analysis to the well-known Sugawara construction.

In section \ref{sect_weylbms}, we investigated the deformations of $\mathfrak{witt}\oplus\mathfrak{H}_3$ \eqref{Heisenberg-direct-sum-algebra-3d} with and without allowing central extensions. In this case, the analysis was technically more involved and we focused on identifying relevant symmetry algebras and the connection of $\mathfrak{witt}\oplus\mathfrak{H}_3$ with the asymptotic symmetry algebra $\mathfrak{bmsw}_3$ \eqref{eq:weylbms3}.

\subsection*{Further discussions on main results}

Let us summarize and discuss further our main results: 
\begin{itemize}
    \item Through deformation of $\mathfrak{H}_3$, we obtained various well-known asymptotic and near horizon symmetry algebras which support the aforementioned intuition that deformations connect different symmetry algebras. These algebras can be organized in two three-parametric deformation families of algebras, the mother algebras $\widehat{W}_{\nu}(0,b)$ \eqref{Deform-G,F,BarG} and $\mathcal{H}_3(\alpha,\nu,\eta)$ \eqref{Deform-G,F,BarF,BarG,TildeF,TildeG}, together with their corresponding central extensions. Their deformation/contraction relationships were collected in figure \ref{Fig.1} and discussed in section \ref{sect_summaryalgebras}. 
    \item In particular, we showed that the two near horizon algebras $\mathfrak{H}_3$ and $W(0,b)$, obtained by imposing various boundary conditions introduced in \cite{Grumiller:2019fmp}, are related through deformation procedure. In addition, we found that $\mathfrak{H}_3$ can be deformed into two copies of Virasoro algebras $\mathfrak{vir}\oplus\mathfrak{vir}$ which is obtained as near horizon and asymptotic symmetry algebras \cite{Adami:2020ugu, Brown:1986nw}. This connection has been previously discovered via a Sugawara construction \cite{Afshar:2016kjj}. 
    
    Furthermore, it has been shown that the algebra $\widehat{W}(0,0)$ can be obtained as asymptotic symmetry algebra of asymptotically AdS$_{3}$ spacetimes \cite{Compere:2013bya} and $\widehat{W}(0,-1)$, also known as $\widehat{\mathfrak{bms}}_{3}$, emerged as asymptotic symmetry algebra of 3-dimensional flat spacetimes \cite{Barnich:2006av}. Recently, $\widehat{W}(0,1)$ has been encountered as asymptotic symmetry algebra in Minkowskian Jackiw-Teitelboim gravity \cite{Afshar:2021qvi}. All of these algebras have been shown to be connected via deformation to $\mathfrak{H}_{3}$ in this work.
    
    In this way, we have shown for explicit examples that, although there are various choices of boundary conditions in the gravitational context, their corresponding symmetry algebras are connected through deformation/contraction procedure.

    \item We further noticed in section \ref{sec_sugawara} that, although some of the new algebras obtained through deformation can  also be found by means of Sugawara constructions, this is not a general feature. Our results provide evidence that the deformation procedure reaches more algebras. Specifically, we gather evidence supporting the statement that those algebras belonging to the family $\widehat{W}_{\nu}(0,b)$ cannot be obtained via a Sugawara construction for arbitrary $b$, while those within $\mathcal{H}_3(\alpha,\nu,\eta)$ can be reached using this procedure.
    
    \item By deforming $\mathfrak{witt}\oplus\mathfrak{H}_3$, we found, among other algebras, the direct sum of three Virasoro algebras or the direct sum of centrally extended $W(0,b)$ and the Virasoro algebra which have been identified as asymptotic symmetry algebras of Maxwell Chern-Simons and AdS-Lorentz Chern-Simons gravity theories in \cite{Adami:2020xkm,Concha:2018jjj}. In addition, the algebra $\mathfrak{vir}\oplus\mathfrak{vir}\oplus\mathfrak{u}(1)$, which has been realized as asymptotic symmetry algebra of three-dimensional Anti-de-Sitter spacetimes when extra Weyl symmetry is considered \cite{Alessio:2020ioh,Geiller:2021vpg}, showed up as a deformation of $\mathfrak{witt}\oplus\mathfrak{H}_3$. Remarkably, all the deformations we derived come uniquely from deforming the Heisenberg part of the algebra, which leads us to speculate that, for two infinite-dimensional algebras $\mathfrak{g_1}\oplus\mathfrak{g_2}$, in which $\mathfrak{g}_1$ is rigid, all the deformations come from the $\mathfrak{g_2}$ part. Examples supporting this conjecture are the deformations obtained within this work for $\mathfrak{witt}\oplus\mathfrak{H}_3$ and the fact that $\mathfrak{vir}\oplus\mathfrak{vir}$ is rigid \cite{FarahmandParsa:2018ojt}. 
    
    \item Our analysis of deformations discarded the possibility that $\mathfrak{witt}\oplus\mathfrak{H}_3$ could be connected through direct continuous deformation to the $\mathfrak{bmsw}_3$ algebra \eqref{eq:weylbms3} and to the null boundary algebra given by (3.34) in \cite{Adami:2020ugu}. Contrarily, we showed that the relation with $\mathfrak{bmsw}_3$ is more involved and consists of a contraction followed by a double deformation procedure. Instead, one can start from a twisted double Heisenberg algebra \eqref{eq:toWeyl-BMS1} and reach $\mathfrak{witt}\oplus\mathfrak{H}_3$, as well as $\mathfrak{bmsw}_3$ and (3.34) in \cite{Adami:2020ugu}.
\end{itemize}

\subsection*{Future research}

Finally, we list some especially interesting research directions. 
\begin{itemize}
    \item As mentioned before, we expect that further investigations on the algebraic relationships between symmetry algebras eventually lead to the development of a mathematical framework, which could help to improve the selection of boundary conditions and shed light on the microstructure of black holes. Further research in this area is highly encouraged.
    
    \item In the fundamental slicing, the boundary algebras are given by $\mathfrak{H}_3\oplus\text{Diff}(\mathcal{N}_v)$, where $\mathcal{N}_v$ is the codimension two compact spacelike surface at constant $v$ \cite{Adami:2020ugu}. In the phenomenologically relevant four-dimensional case with spherical boundaries, this algebra becomes $\mathfrak{H}_3\oplus\text{Diff}(S^2)$. In a previous work \cite{Enriquez-Rojo:2021rtv}, we have investigated the deformations of $\text{Diff}(S^2)$ and, in this work, those of $\mathfrak{H}_3$ have been explored. A very appealing research line which might provide information about the microstructure of four-dimensional black holes is to merge and complement both studies. 
    
    \item Some of the new algebras we obtained as a result of the deformation procedure, such as \eqref{MostDeform-JJ}, \eqref{Deform-JP-barG}, \eqref{Deform-G,BarG} and \eqref{Virasoropluscurrent}, have not yet appeared in a boundary symmetry analysis. It would be interesting to explore whether these algebras can actually arise under new choices of boundary conditions. Following the approach of \cite{Adami:2021kvx}, one could investigate their corresponding thermodynamical interpretation and help to concretize the potential relation between the deformation parameters and thermodynamical or other physical quantities.
    
    \item In section \ref{sect_weylbms}, we analyzed in depth the deformation/contraction relationship between the algebras $\mathfrak{witt}\oplus\mathfrak{H}_3$ and $\mathfrak{bmsw}_3$. Indeed, this is the $n=1$ version of the relation between the near horizon symmetry algebra $\mathfrak{H}_{3} \oplus \text{Diff}(S^n)$ and the asymptotic symmetry algebra $\mathfrak{bmsw}_{n+2}$ \footnote{The algebra $\mathfrak{bmsw}_{4}$ has been recently proposed to describe asymptotically flat spacetimes at $\mathcal{I}^+$ when Weyl scalings are allowed $\delta\sqrt{g_{S^2}}\neq0$ \cite{Freidel:2021fxf}. The general algebras $\mathfrak{bmsw}_{n+2}$ have been previously obtained in \cite{Adami:2020amw,Adami:2020ugu,Adami:2021sko}, where closely related symmetry algebras considering the inclusion of Weyl scaling are also discussed.}. Concretely, although the case $n=2$ is beyond the scope of this work, we would like to highlight two important facts. Firstly, the rigidity of $\text{Diff}(S^2)$ under linear deformations may play a decisive role \cite{Enriquez-Rojo:2021rtv} \footnote{Due to the fact that $\text{Diff}(S^2)$ is rigid, $\mathfrak{H}_3\oplus\text{Diff}(S^2)$ can be also regarded as an example to test our postulate that for two infinite-dimensional algebras $\mathfrak{g_1}\oplus\mathfrak{g_2}$, in which $\mathfrak{g}_1$ is rigid, all the deformations come from the $\mathfrak{g_2}$ part.}. Secondly, it is also known that, in contrast to $\mathfrak{witt}$, $\text{Diff}(S^2)$ does not admit central extensions \cite{Bars:1988uj}. 
    
    \item From a holographic perspective, interpolation between the near horizon and asymptotic region may be interpreted as an RG flow in the dual field theory side. This idea was discussed in \cite{FarahmandParsa:2018ojt,Safari:2020pje} and supported, at the level of algebras, by the results of this work. For example, in the context of the $\text{AdS}_3$/$\text{CFT}_2$ correspondence, the near horizon and asymptotic regions of $\text{AdS}_3$ would correspond to IR and UV regions in its dual $\text{CFT}_2$. At the same time, imposing specific boundary conditions, the near horizon symmetry algebra is given by $\mathfrak{H}_3$ while the asymptotic symmetry algebra corresponds to $\mathfrak{vir}\oplus\mathfrak{vir}$. In this work, we showed that these two algebras are related to each other through deformation/contraction procedure. As a consequence, deformation/contraction might be related to an RG flow between UV and IR fixed points of the dual $\text{CFT}_2$. Further research in this area is very promising.
    
    \item In recent works, Lie algebroids \cite{Compere:2019bua} as well as several non-linear algebras \cite{Grumiller:2019fmp,Fuentealba:2020zkf,Fuentealba:2021yvo} (e.g. \eqref{Poisson-CR1} in $D \geq 4$) have been uncovered as asymptotic and near-horizon symmetry algebras. A finite dimensional example of such non-linear algebras is $\kappa$-Poincaré \cite{Amelino-Camelia:2002cqb, Amelino-Camelia:2010lsq} in the context of Doubly Special Relativity (DSR). The study of deformations we developed in this work can only relate Lie algebras to other Lie algebras and, therefore, the aforementioned examples cannot be obtained using our techniques. As a consequence, it would be interesting to formalize a broader notion of non-linear deformations which can relate Lie algebras at the level of enveloping algebras. 
    
    \item Interpretation of deformation/contraction of Lie algebras at the level of action has been pursued in \cite{Kraus:2021cwf,Rodriguez:2021tcz,He:2021bhj}. It would be very interesting to extend the corresponding analysis to the deformations encountered along this work.
\end{itemize}

To conclude, considering the ubiquity and relevance of the Heisenberg algebra in Physics, we expect that the content and results within this work can have yet unknown applications and implications beyond the context of asymptotic and boundary symmetries.

\section*{Acknowledgements}

We would like to thank M.~M.~Sheikh-Jabbari for initial collaboration, very fruitful discussions and helpful feedback on the manuscript, H.~Afshar for useful comments and I.~Kharag for proofreading this paper. The work of MER was funded by the Excellence Cluster Origins of the DFG under Germany’s Excellence Strategy EXC-2094 390783311.

\appendix

\section{Spacetime structure near generic null surfaces}
\label{app_spacetimeGNS}

Heisenberg algebras have a ubiquitous appearance in the boundary/surface charge algebras, especially those computed on a null surface or on the horizon \cite{Afshar:2016kjj, Afshar:2016wfy,Grumiller:2019fmp,Adami:2020ugu,Adami:2021kvx,Adami:2021nnf}. Let us parametrize the null surface ${\cal N}$ by $v, x^A, A=1,2,\dots, D-2$ and assume it has the topology of $\mathbb{R}\ltimes {\cal N}_v$ (the topology of ${\cal N}_v$ is not fixed) where ${\cal N}_v$ is the codimension two compact spacelike surface and denoted constant $v$ slice on ${\cal N}$. It has been argued in \cite{Adami:2020ugu,Adami:2021kvx,Adami:2021nnf} that the algebra of surface charges depends on the phase space slicing used. In particular, for two such slicings the maximal boundary algebra for a null surface in a $D$ dimensional pure Einstein gravity takes the following forms.

\paragraph{Null boundary algebra in the ``thermodynamic slicing''.} 
\begin{subequations}\label{Thermodynamic-slicing-algebra}
    \begin{align}
   \{{\cal T}(v, x), {\cal T}(v', x')\} &= \left({\cal T}(v,x)\partial_{v'}-\mathcal{T}(v',x')\partial_v\right)\delta(v-v')\delta^{D-2}(x-x'), \label{TT-commutator}\\
       \{{\cal W}(v, x), {\cal W}(v', x')\} &= 0, \\ 
      \{{\cal J}_A(v, x), {\cal J}_B(v', x')\}&= \left(\mathcal{J}_{A}(v,x')\partial_{B}-\mathcal{J}_{B}(v,x)\partial'_{A}\right)\delta^{D-2}(x-x')\delta(v-v'),  \\
      \{ {\cal T}(v,x), {\cal W}(v',x')\} &=  {\cal W}(v,x) \partial_v\delta(v-v')\delta^{D-2}(x-x'),\\
       \{{\cal T}(v,x), {\cal J}_A(v',x')\} &= \left(-{\cal T}(v,x)\partial_A + {\cal J}_A(v,x)\partial_v\right)\delta^{D-2}(x-x') \delta(v-v'), \\
        \{{\cal W}(v,x), {\cal J}_A(v',x')\} &=  {\cal W}(v,x)\partial_A \delta^{D-2}(x-x') \delta(v-v').
     \end{align}
\end{subequations}
The above algebra is WDiff(${\cal N}$), that is Diff(${\cal N}$), generated by ${\cal T}(v,x)$ and ${\cal J}_A(v,x)$, extended by Weyl scaling ${\cal W}(v,x)$. Since topologically ${\cal N}\sim \mathbb{R}\ltimes {\cal N}_v$, the ${\cal T}$ generator forms a Witt algebra at each slice of ${\cal N}_v$ \eqref{TT-commutator}. Moreover, the subalgebra spanned by ${\cal T}, {\cal W}$
is in fact $U(1)$ Kac-Moody algebra (at each slice of ${\cal N}_v$), which can be deformed into BMS$_3$ algebra. Therefore, the algebra may also be viewed as (BMS$_3$)$_{_{{\cal N}_v}}\ltimes$ Diff(${\cal N}_v$). 

\paragraph{Null boundary algebra in ``Heisenberg-Direct sum slicing''.} 

The algebra of charges in this slicing takes the form of Heisenberg $\oplus$ Diff(${\cal N}_v$), i.e.
\begin{subequations}\label{Heisenberg-direct-sum-algebra}
    \begin{align}
        &\{{\cal Q}(v,x),{\cal Q}(v,x')\}=\{\mathcal{P}(v,x),\mathcal{P}(v,x')\}=0,\\
        &\{{\cal Q}(v,x),\mathcal{P}(v,x')\}=\delta^{D-2}\left(x-x'\right),\\
        &\{\mathcal{J}_A(v,x),{\cal Q}(v,x')\}=\{\mathcal{J}_A(v,x),\mathcal{P}(v,x')\}=0,\\
        &\{\mathcal{J}_A(v,x),\mathcal{J}_B(v,x')\}=\left(\mathcal{J}_{A}(v,x')\partial_{B}-\mathcal{J}_{B}(v,x)\partial'_{A}\right)\delta^{D-2}\left(x-x'\right).
    \end{align}
\end{subequations}
The Diff(${\cal N}_v$) part may also admit central extensions. In particular, for the $D=3$ case in the Topologically Massive Gravity (TMG) theories, it has been shown that there exists such a central extension \cite{Adami:2021sko}. Moreover, in special cases where ${\cal N}_v$ has toroidal topology, Diff(${\cal N}_v$) may be replaced by other Heisenberg algebras.

As we see, while the charges have arbitrary $v$ dependence, the RHS of the commutators, the structure constants of the algebra, are $v$ independent. That means that we have the same algebra at any constant $v$ slice. This is to be contrasted with the case in \eqref{Thermodynamic-slicing-algebra}.

\paragraph{Non-expanding null surface algebra.} The above charges and algebra hold for generic null surfaces. However, in the interesting and important special case of non-expanding surface (when the expansion parameter of the surface vanishes), one loses the ${\cal P}$ tower of charge and, importantly, the $v$ dependence of the $\tilde{{\cal J}_A}, {\cal Q}$ is fixed by gravity equations of motion. In this case, and in an appropriate slicing of the solution phase space, we remain with algebra
\begin{subequations}\label{BMS-like-non-expanding-algebra}
    \begin{align}
        &\{{\cal Q}(x),{\cal Q}(x')\}=0,\\
        &\{\tilde{{\cal J}_A}(x),{\cal Q}(x')\}=-{\cal Q}(x)\partial_{A}'\delta^{D-2}(x-x'),\\
        &\{\tilde{{\cal J}_A}(x),\tilde{{\cal J}_B}(x')\}=\left(\tilde{{\cal J}_A}(x')\partial_{B}-\tilde{{\cal J}_B}(x)\partial_{A}'\right)\delta^{D-2}(x-x').
    \end{align}
\end{subequations}
The above algebra is WDiff(${\cal N}_v$), where the Diff(${\cal N}_v$) part is generated by ${\cal J}_A(x)$ and the Weyl scaling by ${\cal Q}(x)$. 

\paragraph{Non-expanding null surface algebra in Heisenberg slicing.}
Upon redefinition of the generator $\tilde{{\cal J}_A}\to \tilde{{\cal J}_A}/{\cal Q}$, \eqref{BMS-like-non-expanding-algebra} takes the form
\begin{equation}\label{Poisson-CR1}
    \begin{split}
    &\{{\cal Q}(x),{\cal Q}(y)\}=0,\\
    &\{\tilde{{\cal J}_A}(x),{\cal Q}(y)\}=\frac{\partial}{\partial x^{A}}\delta(x-y),\\
        &\{\tilde{{\cal J}_A}(x),\tilde{{\cal J}_B}(y)\}={\cal Q}^{-1}(x)\tilde{F}_{BA}(x)\delta(x-y),
    \end{split}
\end{equation}
where $\tilde{F}_{AB}=\partial_{A}\tilde{{\cal J}_{B}}-\partial_{B}\tilde{{\cal J}_{A}}$. The $\tilde{{\cal J}_A}(x)$ charge is a one-form on ${\cal N}_v$ and may be decomposed into exact and coexact parts using Hodge decomposition:
\be\label{Hodge-decomposition}
\tilde{{\cal J}_{A}}=8\pi G\partial_{A}{\Pi}+\nabla^B {\cal J}_{AB} \ ,
\ee
where ${\cal J}_{AB}$ is a two-form and $\nabla^B$ is the covariant derivative on ${\cal N}_v$. $\tilde{F}_{AB}$ is therefore only depending on the coexact part ${\cal J}_{AB}$. One then observes that 
\begin{equation}\label{Heisenberg-JHA}
 \begin{split}
     & \{{\cal Q}(x),\Pi(y)\}=\frac{1}{8\pi G}\delta(x-y)\\
&\{{\cal Q}(x),{\cal J}_{AB}\}=0 \ .
\end{split}
\end{equation}
One may then work out the other commutators of the algebra. Notoriously, we have a Heisenberg part in the algebra again. A crucial comment is in order here. One may readily observe that the ``entropy'' $S\propto \int {\cal Q}$ commutes with all $\tilde{{\cal J}_{A}}(x)$, i.e. $\tilde{{\cal J}_{A}}$ are the charges associated with area-preserving diffeomorphisms. This feature, however, is not respected by \eqref{Heisenberg-JHA}; zero mode of $\Pi$, $\Pi_0$ does not commute with $S$. This is due to the fact that $\Pi_0$ does not appear in \eqref{Hodge-decomposition} and $\Pi_0$ is not a part of the original algebra \eqref{Poisson-CR1}.

We stress that the maximal null boundary algebras are not limited to the two cases \eqref{Thermodynamic-slicing-algebra} and \eqref{Heisenberg-direct-sum-algebra} and there are many more such algebras obtained through change of slicing on the null boundary phase space. Nonetheless, it is believed that all these algebras can be obtained from deformations of these two (for further discussions, look at \cite{Adami:2020ugu,Adami:2021kvx,Adami:2021nnf}). The deformations obtained throughout this work provide further evidence in this regard.

\subsection{Three-dimensional case}

So far, we discussed four different algebras, \eqref{Thermodynamic-slicing-algebra}, \eqref{Heisenberg-direct-sum-algebra}, \eqref{BMS-like-non-expanding-algebra} and \eqref{Poisson-CR1}.  In the three-dimensional case they take simpler forms \cite{Adami:2020ugu,Adami:2021sko}. Let us discuss each of these three cases separately. In all cases we ``quantize'' the algebra by replacing Poisson bracket with Dirac bracket as $\{,\}\rightarrow i[,]$.

\paragraph{$\text{WDiff}(C_2)$ algebra.} In this case, the generators of \eqref{Thermodynamic-slicing-algebra} are functions over a two-dimensional cylinder $C_2$ spanned by $v, x$ ($x\sim x+2\pi$). One may Fourier expand in $x$ and, assuming meromorphicity, one can also Laurent expand in $v$, 
\begin{equation}\label{Fourier-Laurent-3d}
    \mathcal{T}(v,x)=\sum^{+\infty}_{-\infty}\mathcal{P}_{mn} v^m e^{inx}, \qquad {\cal W}(v,x)=\sum^{+\infty}_{-\infty}\mathcal{Q}_{mn} v^m e^{inx},\qquad {\cal J}(v,x)=\sum^{+\infty}_{-\infty}\mathcal{J}_{mn} v^m e^{inx}.
\end{equation}
Then \eqref{Thermodynamic-slicing-algebra}  takes the form
\begin{subequations}\label{WDiffC2-algebra-3d}
    \begin{align}
         &[{\cal P}_{m,n},{\cal P}_{p,q}]=(m-n) {\cal P}_{m+p, n+q},\\
        &[\mathcal{Q}_{m,n},{\cal P}_{p,q}]=-(m+1){\cal Q}_{m+p,n+q},\\
         &[{\cal Q}_{m,n},{\cal Q}_{p,q}]=0,\\
          &[{\cal Q}_{m,n},{\cal J}_{p,q}]= -n\mathcal{Q}_{m+p+1,n+q}\\
          &[{\cal P}_{m,n},{\cal J}_{p,q}]= -n\mathcal{P}_{m+p+1,n+q}+ (p+1)\mathcal{J}_{m+p,n+q}-n(p+1)\mathcal{Q}_{m+p,n+q},\\
        &[\mathcal{J}_{m,n},\mathcal{J}_{p,q}]= (n-q)\mathcal{J}_{m+p+1,n+q}+\frac{c}{12}n^3 \delta_{n+q,0}\delta_{m+p+1,0},
    \end{align}
\end{subequations}
where we have included the possible central extension in the ``superrotation'' part. This algebra has many interesting subalgebras, e.g. one may restrict to ${\cal X}_{m,0}$ or ${\cal X}_{-1,m}$ sectors, which both constitute a Witt algebra extended with two Abelian currents.

\paragraph{Heisenberg $\oplus$ Diff($S^1$).} Since the structure constants are independent of $v$, for the sake of simplicity one may suppress the $v$-dependence of the charges in \eqref{Heisenberg-direct-sum-algebra}. Moreover, one may Fourier expand the three towers of charges:
\begin{equation}\label{Fourier-3d}
    \mathcal{J}(x)=\sum^{+\infty}_{-\infty}\mathcal{J}_{m} e^{imx}, \qquad {\cal Q}(y)=\sum^{+\infty}_{-\infty}\mathcal{Q}_{n} e^{inx},\qquad {\cal P}(x)=\sum^{+\infty}_{-\infty}\mathcal{P}_{n} e^{inx}.
\end{equation}
The algebra \eqref{Heisenberg-direct-sum-algebra} then takes the form 
\begin{subequations}
    \begin{align}
        &[{\cal Q}_m,{\cal Q}_n]=[\mathcal{P}_m,\mathcal{P}_n]=0,\\
        &[{\cal Q}_m,\mathcal{P}_n]= i\hbar \delta_{m+n,0},\\
        &[\mathcal{J}_m,{\cal Q}_n]=[\mathcal{J}_m,\mathcal{P}_n]=0,\\
        &[\mathcal{J}_m,\mathcal{J}_n]= (m-n)\mathcal{J}_{m+n} +\frac{c}{12}m^3 \delta_{m+n,0},
    \end{align}
\end{subequations}
that is \eqref{Heisenberg-direct-sum-algebra-3d}. Here we have also added a central term $c$ which arises in the TMG case \cite{Adami:2021sko}.

\paragraph{Virasoro-Kac-Moody algebra.} The algebra \eqref{BMS-like-non-expanding-algebra} in terms of Fourier modes, and after inclusion of the central term takes the form
\begin{subequations}
    \begin{align}
         &[{\cal Q}_m,{\cal Q}_n]={\tilde{c}}\,m\, \delta_{m+n,0},\\
        &[\mathcal{J}_m,{\cal Q}_n]=- n{\cal Q}_{m+n}+{\bar{c}}\,m^2\, \delta_{m+n,0},\\
        &[\mathcal{J}_m,\mathcal{J}_n]= (m-n)\mathcal{J}_{m+n}+\frac{c}{12}\,m^3 \delta_{m+n,0},
    \end{align}
\end{subequations}
which corresponds to \eqref{BMS-like-non-expanding-algebra-3d}.

\paragraph{Heisenberg-like algebra.} In 3d case $F_{AB}=0$ and hence, \eqref{Poisson-CR1} takes the form
\begin{equation}
    \begin{split}
    &[\mathcal{J}_{m},\mathcal{J}_{n}]=0,\\
    &[\mathcal{J}_{m},\mathcal{P}_{n}]=i \hbar \,m\delta_{m+n,0},\\
        &[\mathcal{P}_{m},\mathcal{P}_{n}]=0,
    \end{split}
\end{equation}
obtaining \eqref{Dirac-3d}. Using a redefinition $\mathcal{J}_{m}\rightarrow m\,\mathcal{J}_{m}$, we obtain the infinite dimensional Heisenberg algebra \cite{Afshar:2016kjj, Afshar:2016wfy} denoted by $\mathfrak{H}_3$.

\section{Deformation theory of Lie algebras}
\label{app_deformations}

In this appendix, we briefly review the concept of deformation of Lie algebras, their relation to the cohomology of Lie algebras and define the terminology used in the main body of the text, indicating the precise deformations we analyzed in this work. We refer the reader to \cite{Safari:2020pje, fuks2012cohomology} for more details. 

A {\it{deformation}} of a certain Lie algebra $\mathfrak{g}$ is a modification of its structure constants. Some of such deformations could just be a change of basis, these are called trivial deformations. Non-trivial deformations modify/deform a Lie algebra $\mathfrak{g}$ to another Lie algebra with the same vector space structure \footnote{In the case of finite dimensional Lie algebras the latter implies that deformation does not change the dimension of the algebra.}. 
The concept of deformation of rings and algebras was first introduced in a series of papers by Gerstenhaber \cite{gerstenhaber1964deformation, gerstenhaber1966deformation, gerstenhaber1968deformation, gerstenhaber1974deformation} and by Nijenhuis and Richardson for Lie algebras in \cite{nijenhuis1967deformations}. A Lie algebra $\mathfrak{g}$ is called {\it{rigid}} or stable if it does not admit any deformation or, equivalently, if the algebras obtained from it via deformation $\mathfrak{g}_{\varepsilon}$, $\varepsilon$ being the possible deformation parameters, are isomorphic to the initial algebra $\mathfrak{g}$. Conversely, given an algebra $\mathfrak{g}$ one can take the limit $\varepsilon \to 0$ and obtain $\mathfrak{g}_{0}$. This procedure is known as {\it{contraction}} of Lie algebras \footnote{Let us note that the deformation/contraction procedure differs with the semigroup expansion method, in which one can preserve the contraction parameter and use it as an
expansion parameter rather than just approaching zero \cite{deAzcarraga:2007et, Khasanov:2011jr}. This procedure leads to an extension of the contracted algebra. For instance, the semigroup expansion of a finite algebra at each level in the expansion parameter leads to a larger algebra with more generators.}. The contraction and deformation are hence inverse of each other.

\subsection{Formal deformation of Lie algebras}
We denote by $(\mathfrak{g},[,])$ a Lie algebra in which $\mathfrak{g}$ is a vector space over a field $\mathbb{F}$ with characteristic zero (e.g. $\mathbb{R}$) equipped with a Lie bracket, that is a bilinear and antisymmetric product function 
\begin{equation}
    [,]: \mathfrak{g}\times \mathfrak{g}\longrightarrow \mathfrak{g} \ ,
\end{equation}
which must also satisfy the Jacobi identities
\begin{equation}
    [g_{i},[g_{j},g_{k}]]+[g_{j},[g_{k},g_{i}]]+[g_{k},[g_{i},g_{j}]]=0,\qquad \forall g_{i} \in \mathfrak{g}. \label{GJ}
\end{equation}

{A {\it formal one-parameter deformation} of a Lie algebra ($\mathfrak{g},{[,]}_{0}$), abbreviated as $\mathfrak{g}$, is defined as a skew symmetric bilinear map $\mathfrak{g}\times\mathfrak{g}\rightarrow\mathfrak{g}[[\varepsilon]]$ which satisfies the Jacobi identities to all orders of $\varepsilon$, where $\mathfrak{g}[[\varepsilon]]$ is the space of formal power series in $\varepsilon$ with coefficients in $\mathfrak{g}$ \cite{fox1993introduction}.} This means that the commutation relations of $\mathfrak{g}$ are modified as follows:
 \begin{equation}\label{eq:defdefor}
     {[g_{i},g_{j}]}_\varepsilon:=\Psi(g_{i},g_{j};\varepsilon)=\Psi(g_{i},g_{j};\varepsilon=0)+{{\psi }_1(g_{i},g_{j})}\varepsilon+{{\psi }_2(g_{i},g_{j})}\varepsilon^{2}+... \ ,
 \end{equation}
where $ \Psi(g_{i},g_{j};\varepsilon=0)={[g_{i},g_{j}]}_0$, $g_{i}$ and $g_{j}$ are basis elements of $\mathfrak{g}$, $\varepsilon \in \mathbb{F}$ (e.g. $\mathbb{R}$) is the {\it{deformation parameter}} and ${\psi }_i:\ \mathfrak{g}\times \mathfrak{g}\longrightarrow \mathfrak{g}$ are bilinear antisymmetric functions, the so-called 2-\textit{cochains}. 
For every $\varepsilon$, the new Lie algebra ($\mathfrak{g},{[,]}_{\varepsilon}$) should satisfy the Jacobi identity
\be
[g_{i},[g_{j},g_{k}]_{\varepsilon}]_{\varepsilon}+\text{cyclic permutations of}\ (g_{i},g_{j},g_{k}) =0 \label{Jepsilon}
\ee
 order by order in $\varepsilon$, which leads to an infinite sequence of equations among $\psi_{i}$.    
 
 For small $\varepsilon$ the leading deformation is given by ${\psi }_{1}(g_{i},g_{j})$-term and the associated Jacobi identities lead to
\be\label{coe}
  {[g_{i},{\psi }_1(g_{j},g_{k})]}_0+{\psi }_1(g_{i},{[g_{j},g_{k}]}_0)+\text{cyclic permutations of}\ (g_{i},g_{j},g_{k}) =0.
\ee
This relation is known as the $2-$cocycle condition. Its solution, the  $2-$cocycle ${\psi }_1$,  specifies an {\it infinitesimal} deformation of Lie algebra $\mathfrak{g}$. The Jacobi identity for higher orders of $\varepsilon$ should also be checked as integrability conditions of ${\psi }_1$ and may lead to obstructions, which will be discussed later in this appendix. From now on, we denote the deformed algebra ($\mathfrak{g},{[,]}_{\varepsilon}$) simply by $\mathfrak{g}_{\varepsilon}$. 

One can readily check that the relation
\begin{equation}
    {\psi }_1(g_{i},g_{j})={{\varphi }_1([g_{i},g_{j}])}_0-{[{\varphi }_1(g_{i}),g_{j}]}_0-{[g_{i},{\varphi }_1(g_{j})]}_0,\label{cob}
\end{equation}
 satisfies the 2-cocycle condition \eqref{coe}. In fact \eqref{cob} shows that ${\psi}_1$ is a 2-coboundary if ${\varphi }_1$ is a 1-cochain. When ${\psi }_1$ is a 2-coboundary the deformation \eqref{eq:defdefor} is called trivial, meaning that the deformation is just a redefinition of the basis elements.

\subsection{Relation of deformation theory and cohomology of Lie algebras}\label{sec:3.3}

We start with the definition of the Chevalley-Eilenberg complex and differential.
A vector space $\mathbb{V}$ is called a $\mathfrak{g}$-module if {there exists}
a bilinear map $\omega:\ \mathfrak{g}\times \mathbb{V}\longrightarrow \mathbb{V}$  
for all $x\in \mathbb{V}$ and $g_{1},g_{2} \in \mathfrak{g}$ with the property $\omega([g_{1},g_{2}],x)=\omega(g_{1},\omega(g_{2},x))-\omega(g_{2},\omega(g_{1},x))$  \cite{fuks2012cohomology}. In this setting, the Jacobi identities of the Lie bracket imply that a Lie algebra $\mathfrak{g}$ with the adjoint action is a $\mathfrak{g}$-module. 
{A $p$-cochain $\psi$ is a $\mathbb{V}$-valued (as $\mathfrak{g}$-module),  
 bilinear and completely antisymmetric function which is defined as:
\begin{align*}
    \psi: \underbrace{\mathfrak{g}\times\cdots\times\mathfrak{g}}_{p \, \ times}&\longrightarrow \mathbb{V}\\
    (g_1,\cdots ,g_p)&\longmapsto \psi(g_1,\cdots ,g_p);\,\,\,\,\,\ 0 \leq p \leq dim(\mathfrak{g}).
\end{align*}
Suppose ${\mathcal{C}}^p(\mathfrak{g};\mathbb{V})$ is the space of $\mathbb{V}$-valued $p$-cochains on $\mathfrak{g}$. One can then define the cochain complex ${\mathcal{C}}^*(\mathfrak{g};\mathbb{V})=\oplus^{dim(\mathfrak{g})}_{p=o}{\mathcal{C}}^p(\mathfrak{g};\mathbb{V})$, which is known as the \textbf{\textit{Chevalley-Eilenberg complex}}.

The \textbf{\textit{Chevalley-Eilenberg differential}} or equivalently \textbf{\textit{coboundary operator}} ``$d$'' is a linear map defined as \cite{ChevalleyEilenberg, MR0054581}
\begin{align*}
d_{p}:\ {\mathcal{C}}^{p} (\mathfrak{g} ;\mathfrak{g})&\longrightarrow{\mathcal{C}}^{p+1}\left(\mathfrak{g};\mathfrak{g}\right),\\
     \psi &\longmapsto d_{\mathrm{p}}\psi,
\end{align*}
and the $p+1$-cochain $d_{{p}}\psi$ is given by:
   \begin{align}\label{p-cochain}
     (d_{p}\psi )\left(g_0,\dots ,g_{p}\right)\equiv&\sum_{0\leq {i}{<}{j}\leq {p}}{({{1}})^{{i}{+}{j}{-}{1}}} {\psi }\left(\left[{{g}}_{{i}},{{g}}_{{j}}\right],{{g}}_{{0}},{\dots },\widehat{{{g}}_{{i}}},{\dots },\widehat{{{g}}_{{j}}},{\dots },{{g}}_{{p}{+1}}\right)\cr {+}&
    \sum_{{1}\leq {i}\leq {p}{+1}}{({{-}{1}})^{{i}}}\left[{{g}}_{{i}},{\psi }\left({{g}}_{0},{\dots },\widehat{{{g}}_{{i}}}, {\dots },{{g}}_{p}\right)\right],
\end{align}
where the hat denotes omission. One can check that $d_{p}\circ d_{{p}{-}{1}}{=0.\ }$
A $p$-cochain ${\psi}$  is called \textbf{\textit{$p$-cocycle}} if  ${d}_p{\psi}=0$, and \textbf{\textit{$p$-coboundary}} if $\ {\psi}={d }_{p-1}{\varphi}$.

By means of the property $d_{{p}}\circ d_{{p}{-}{1}}{=0\ }$, one concludes that every $p$-coboundary is also a $p$-cocycle. With this definition one can check that 2-cocycle condition \eqref{coe} is just $d_{2} \psi_{1}=0$, where $\psi$ is a $\mathfrak{g}$-valued $2-$cochain and $d_{2}$  given in \eqref{p-cochain}, and the relation \eqref{cob} is a $2-$coboundary condition $\psi_{1}=d_{1}\varphi_{1}$, where $\varphi_{1}$ is a $\mathfrak{g}$-valued $1-$cochain.

{One defines $Z^{p}(\mathfrak{g};\mathbb{V})$ as a space of $p$-cocycles which is the kernel of the differential $d$ as
\begin{equation}
    Z^{p}(\mathfrak{g};\mathbb{V})=\{\psi\in\mathcal{C}^p(\mathfrak{g};\mathbb{V})|d_{p}\psi=0 \}.
\end{equation}
$Z^{2}(\mathfrak{g};\mathfrak{g})$ is hence the space of all $\mathfrak{g}-$valued $2-$cocycles which satisfy the relation \eqref{coe}. 
One also define $B^{p}(\mathfrak{g};\mathbb{V})$ as the space of $p$-coboundaries in the following way
\begin{equation}
    B^{p}(\mathfrak{g};\mathbb{V})=\{\psi\in\mathcal{C}^p(\mathfrak{g};\mathbb{V})|\psi=d_{p-1}\varphi \,\,\, \text{for some}\ \ \varphi\ \text{in}\ \  \mathcal{C}^{p-1}(\mathfrak{g};\mathbb{V}) \}.
\end{equation}
$B^{2}(\mathfrak{g};\mathfrak{g})$ is, therefore, the space of all $\mathfrak{g}-$valued $2-$cocycles which are also $2-$coboundaries, meaning that its elements satisfy both relations \eqref{coe} and \eqref{cob}. A $p^{th}\ cohomology$ space of $\mathfrak{g}$ with coefficients in $\mathbb{V}$ is then defined as the quotient of the space of $p$-cocycles $ Z^{p}(\mathfrak{g};\mathbb{V})$ to the space of  $p$-coboundaries $ B^{p}(\mathfrak{g};\mathbb{V})$ as
 \begin{equation}
     {\mathcal{H}}^p(\mathfrak{g};\mathbb{V}):=Z^{p}(\mathfrak{g};\mathbb{V})/B^{p}(\mathfrak{g};\mathbb{V})=\text{Ker}({\ d }_p)/\text{Im}({\ d }_{p-1}).
 \end{equation}
 
 It is worth to point out that isomorphic Lie algebras have the same cohomology spaces and that ${\mathcal{H}}^{2}(\mathfrak{g};\mathfrak{g})$, the second adjoint cohomology, classifies all infinitesimal deformations of algebra $\mathfrak{g}$. Not all infinitesimal deformations integrate to a formal (finite) deformation, there could be obstructions.


\paragraph{Integrability conditions and obstructions.} \label{Integrability}
As pointed out earlier, in order to have a formal deformation \eqref{eq:defdefor}, we need the corresponding non-trivial infinitesimal deformation to be integrable, that is to be valid to all orders in the deformation parameter. To the first few orders in $\varepsilon$, \eqref{Jepsilon}  leads to 
\begin{subequations}\label{infs}
\begin{align}   &[g_{i},[g_{j},g_{k}]_{0}]_{0}+\text{cyclic permutation of}\ (g_{i},g_{j},g_{k})=0 \ , \label{infs-a}\\
    &d_{2}\psi_{1}=0 \ , \label{infs-b}\\
    &d_{2}\psi_{2}=-\frac{1}{2}[\![\psi_{1},\psi_{1}]\!] \ , \label{infs-c}\\
   & d_2\psi_3=-[\![\psi_1,\psi_2]\!] \ , \label{infs-d} \\
   & \cdots \nonumber
\end{align}
\end{subequations}
where we used the definition of the {Chevalley-Eilenberg differential} $d_{2}$ in \eqref{p-cochain} and the double-bracket is the Nijenhuis and Richardson bracket \cite{nijenhuis1967deformations} defined as 
$$\frac{1}{2}[\![\psi_{r},\psi_{s}]\!](g_{i},g_{j},g_{k}):=\psi_{r}(g_{i},\psi_{s}(g_{j},g_{k}))+\text{cyclic permutation of}\ (g_{i},g_{j},g_{k}).
$$
The zeroth order in $\varepsilon$, \eqref{infs-a}, is nothing but the Jacobi relation for the undeformed algebra and is hence satisfied by definition. The second equation \eqref{infs-b} is the $2-$cocycle condition \eqref{coe} for $\psi_{1}$ and its solutions provides nontrivial \emph{infinitesimal} deformations. Eq.\eqref{infs-c} would then guarantee that there are no obstructions in viewing $\psi_1(g_i,g_j)$ as the first order term of a formal deformation $\Psi(g_i,g_j;\eps)$ which admits a power series expansion in $\eps$. Naturally, one should continue the same reasoning in higher orders of $\eps$. For example, it is readily observed that for order $\eps^3$ one should satisfy \eqref{infs-d}. The sequence of relations will stop if there is an obstruction. From a cohomological point of view, one can verify that all obstructions are in the space ${\mathcal{H}}^{3}(\mathfrak{g};\mathfrak{g})$. If ${\mathcal{H}}^{3}(\mathfrak{g};\mathfrak{g})=0$ then there are no obstructions \cite{nijenhuis1967deformations}.

There are three different approaches to check the integrability conditions \footnote{It should be emphasized that other methods than direct calculations are very complex and time consuming, in case they even exist, especially in the case of infinite dimensional algebras. For this reason, we restrict ourselves to the third method within this work.}: 
\begin{itemize}
\item[1.] Sequential method: One can consider the entire infinite sequence of relations \eqref{infs} and directly verify their solutions or probable obstructions. 
\item[2.] ${\cal H}^3$ method: We mentioned that all obstructions are located in $\mathcal{H}^{3}(\mathfrak{g};\mathfrak{g})$. If $\mathcal{H}^{3}(\mathfrak{g};\mathfrak{g})$ vanishes there is no obstruction. For further discussions we refer the reader to \cite{Safari:2020pje}.
\item[3.] Direct method: One may examine if an infinitesimal deformation is indeed formal by promoting the linear (infinitesimal) deformation $\psi_{1}(g_{i},g_{j})$ to $\Psi(g_{i},g_{j};\varepsilon)$ and check whether it satisfies the Jacobi identities or not. If one finds that the linear term in the Taylor expansion of $\Psi(g_{i},g_{j};\varepsilon)$ satisfies the Jacobi identities \eqref{GJ}, one concludes that it is also a formal deformation of algebra.
\end{itemize}

\subsection*{Deformations investigated in this work}

Throughout the text we tackle with infinitesimal and formal deformations of Heisenberg-like algebras. First, we mention that the deformation parameters are considered as a part of the 2-cocycle functions' definition, which are introduced as deformation terms. Looking for infinitesimal deformation is equivalent to keep only the linear term of the functions in the Jacobi identities. After we classify all infinitesimal deformations, which is equivalent to obtain the second adjoint cohomology $\mathcal{H}^{2}(\mathfrak{g};\mathfrak{g})$, we explore which of these infinitesimal deformations can be enhanced to formal deformations using the direct method to evaluate their integrability conditions. Finally, we consider which of these deformations induce a non-trivial deformation. In this way, we determine all formal deformations of Heisenberg-like algebras in sections \ref{sec_defsHeisenbeg} and \ref{sect_weylbms}. 


\section{Analysis of Jacobi identities}
\label{app_Jacobis}

\subsection{Heisenberg algebra}
\label{app_JacobiH3}

In this appendix, we develop the analysis of the Jacobi identities and associated constraints for the ansatz \eqref{eq:moregendefh3}. 

The Jacobi identities $[\mathcal{J}_{m},[\mathcal{J}_{n},\mathcal{J}_{l}]]+\text{cyclic permutations}=0$ lead to the independent relations
\begin{multline}\label{JJJ-Coeff-J}
    (n-l)(m-n-l)F(n,l)F(m,n+l)+(l-m)(n-l-m)F(l,m)F(n,l+m)+ \\
(m-n)(l-m-n)F(m,n)F(l,m+n)+\\
(n-l)G(n,l)\Bar{F}(m,n+l)+(l-m)G(l,m)\Bar{F}(n,l+m)+(m-n)G(m,n)\Bar{F}(l,m+n)=0 \ ,
\end{multline}
 
\begin{multline}\label{JJJ-Coeff-P}
    (n-l)(m-n-l)F(n,l)G(m,n+l)+(l-m)(n-l-m)F(l,m)G(n,l+m)+ \\
(m-n)(l-m-n)F(m,n)G(l,m+n)+\\
(n-l)G(n,l)\Bar{G}(m,n+l)+(l-m)G(l,m)\Bar{G}(n,l+m)+(m-n)G(m,n)\Bar{G}(l,m+n)=0 \ ,
\end{multline}
and 
\begin{multline}\label{FAG-JJ}
(m-n-l)(n-l)F(n,l)A(l+n,m)+(n-l-m)(l-m)F(l,m)A(m+l,n)\\
+(l-m-n)(m-n)F(m,n)A(m+n,l)+\{m(n-l)G(n,l)+n(l-m)G(l,m)+l(m-n)G(m,n)\}\delta_{m+n+l,0}+\\
(n-l)G(n,l)\Bar{A}(m,n+l)+(l-m)G(l,m)\Bar{A}(n,l+m)+(m-n)G(m,n)\Bar{A}(l,m+n)=0 \ .
\end{multline}


The Jacobi identities $[\mathcal{J}_{m},[\mathcal{J}_{n},\mathcal{P}_{l}]]+\text{cyclic permutations}=0$ yield 
\begin{multline}\label{JJP-Coeff-J}
    (m-n-l)\Bar{F}(n,l)F(m,n+l)+\Bar{G}(n,l)\Bar{F}(m,n+l)- (n-l-m)\Bar{F}(m,l)F(n,l+m)-\Bar{G}(m,l)\Bar{F}(n,l+m)+ \\
-(m-n)F(m,n)\Bar{F}(m+n,l)+(m-n)(l-m-n)G(m,n)\tilde{F}(l,m+n)=0 \ ,
\end{multline}
 
\begin{multline}\label{JJP-Coeff-P}
    (m-n-l)\Bar{F}(n,l)G(m,n+l)+\Bar{G}(n,l)\Bar{G}(m,n+l)- (n-l-m)\Bar{F}(m,l)G(n,l+m)-\Bar{G}(m,l)\Bar{G}(n,l+m)+ \\
-(m-n)F(m,n)\Bar{G}(m+n,l)+(m-n)(l-m-n)G(m,n)\tilde{G}(l,m+n)=0 \ ,
\end{multline}
and
\begin{multline}\label{Bar(G-A-F)-A-G-tildeA}
    (m\Bar{G}(n,l)-n\Bar{G}(m,l)-(m+n)(m-n)F(m,n))\delta_{m+n+l,0}\\
    +\Bar{G}(n,l)\Bar{A}(m,n+l)-\Bar{G}(m,l)\Bar{A}(n,l+m)-(m-n)F(m,n)\Bar{A}(m+n,l)\\
    +(m-n-l)\Bar{F}(n,l)A(m,l+n)-(n-m-l)\Bar{F}(m,l)A(n,l+m)+(l-m-n)(m-n)G(m,n)\tilde{A}(l,m+n)=0 \ .
\end{multline}

The Jacobi identities $[\mathcal{P}_{m},[\mathcal{P}_{n},\mathcal{J}_{l}]]+\text{cyclic permutations}=0$ yield 
\begin{multline}\label{PPJ-coeff-J}
    \Bar{F}(l,n)\Bar{F}(n+l,m)-(m-n-l)\Bar{G}(l,n)\tilde{F}(m,n+l)-\Bar{F}(l,m)\Bar{F}(l+m,n)+(n-l-m)\Bar{G}(l,m)\tilde{F}(n,l+m)+ \\
+(m-n)(l-m-n)F(l,m+n)\tilde{F}(m,n)+(m-n)\tilde{G}(m,n)\Bar{F}(l,m+n)=0 \ ,
\end{multline}

\begin{multline}\label{PPJ-Coeff-P}
    -(m-n-l)\Bar{G}(l,n)\tilde{G}(m,n+l)+\Bar{F}(l,n)\Bar{G}(n+l,m)+(n-l-m)\Bar{G}(l,m)\tilde{G}(n,l+m)-\Bar{F}(l,m)\Bar{G}(l+m,n)+ \\
+(m-n)\tilde{G}(m,n)\Bar{G}(l,m+n)+(m-n)(l-m-n)G(l,m+n)\tilde{F}(m,n)=0 \ ,
\end{multline}
and
\begin{multline}\label{Bar(F,A,G)-tilde(A,F,G)}
   ((n+l)\Bar{F}(l,n)-(m+l)\Bar{F}(l,m)+(l)(m-n)\tilde{G}(m,n))\delta_{m+n+l,0}\\
    +\Bar{F}(l,n)\Bar{A}(n+l,m)-\Bar{F}(l,m)\Bar{A}(l+m,n)+(m-n)\tilde{G}(m,n)\Bar{A}(l,m+n)\\
   -(m-l-n)\Bar{G}(l,n)\tilde{A}(m,l+n)+(n-l-m)\Bar{G}(l,m)\tilde{A}(n,m+l)+(l-m-n)(m-n)\tilde{F}(m,n)A(l,m+n)=0 \ .
\end{multline}

Finally, the Jacobi identities $[\mathcal{P}_{m},[\mathcal{P}_{n},\mathcal{P}_{l}]]+\text{cyclic permutations}=0$ lead to
\begin{multline}
    (n-l)(m-n-l)\tilde{G}(n,l)\tilde{G}(m,n+l)+(l-m)(n-l-m)\tilde{G}(l,m)\tilde{G}(n,l+m)+ \\
(m-n)(l-m-n)\tilde{G}(m,n)\tilde{G}(l,m+n)\\
-(n-l)\tilde{F}(n,l)\Bar{G}(n+l,m)-(l-m)\tilde{F}(l,m)\Bar{G}(l+m,n)-(m-n)\tilde{F}(m,n)\Bar{G}(m+n,l)=0 \ ,
\end{multline}

\begin{multline}
    (n-l)(m-n-l)\tilde{G}(n,l)\tilde{F}(m,n+l)+(l-m)(n-l-m)\tilde{G}(l,m)\tilde{F}(n,l+m)+ \\
(m-n)(l-m-n)\tilde{G}(m,n)\tilde{F}(l,m+n)\\
-(n-l)\tilde{F}(n,l)\Bar{F}(n+l,m)-(l-m)\tilde{F}(l,m)\Bar{F}(l+m,n)-(m-n)\tilde{F}(m,n)\Bar{F}(m+n,l)=0 \ ,
\end{multline}
and 
\begin{multline}\label{tilde(F,G,A)-BarA}
\{-(n+l)(n-l)\tilde{F}(n,l)-(l+m)(l-m)\tilde{F}(l,m)-(m+n)(m-n)\tilde{F}(m,n)\}\delta_{m+n+l,0}\\
+(n-l)(m-n-l)\tilde{G}(n,l)\tilde{A}(m,n+l)+(l-m)(n-m-l)\tilde{G}(l,m)\tilde{A}(n,m+l)\\+(m-n)(l-m-n)\tilde{G}(m,n)\tilde{A}(l,m+n)\\
-(n-l)\tilde{F}(n,l)\Bar{A}(n+l,m)-(l-m)\tilde{F}(l,m)\Bar{A}(l+m,n)-(m-n)\tilde{F}(m,n)\Bar{A}(m+n,l)=0 \ .
\end{multline}
First of all, we consider infinitesimal deformations such that we just study the relations which include first order of functions. These relations are given by 
\begin{equation}\label{linear-G}
    (m(n-l)G(n,l)+n(l-m)G(l,m)+l(m-n)G(m,n))\delta_{m+n+l,0}=0 \ ,
\end{equation}
\begin{equation}\label{linear-BarG-F}
    (m\Bar{G}(n,l)-n\Bar{G}(m,l)-(m+n)(m-n)F(m,n))\delta_{m+n+l,0}=0 \ ,
\end{equation}
\begin{equation}
    ((n+l)\Bar{F}(l,n)-(m+l)\Bar{F}(l,m)+(l)(m-n)\tilde{G}(m,n))\delta_{m+n+l,0}=0 \ ,
\end{equation}
\begin{equation}
    (-(n+l)(n-l)\tilde{F}(n,l)-(l+m)(l-m)\tilde{F}(l,m)-(m+n)(m-n)\tilde{F}(m,n))\delta_{m+n+l,0}=0 \ .
\end{equation}
As discussed in the section \ref{sect_H3separate}, the first and last relations have as solutions $G(m,n)=constant$ and $\tilde{F}(m,n)=constant$. 
The second relation is solved by $F(m,n)=constant=\beta$ and $\Bar{G}(m,n)=\alpha m-\beta n$ \footnote{Another solution given by $\Bar{G}(m,n)=\beta m^k  \delta_{m+n,0}$ and $F(m,n)=0$ is possible and will also be discussed.}. 
The same argument is true for $\tilde{G}(m,n)$ and $\Bar{F}(m,n)$ in the third relation. 
So we can recognize four independent infinitesimal deformations by $G(m,n)$, $\{F(m,n),\Bar{G}(m,n)\}$, $\{\Bar{F}(m,n),\tilde{G}(m,n)\}$ and $\tilde{F}(m,n)$ with the mentioned solutions. We should also consider the possible combinations of these deformations.

\subsection{$\mathfrak{witt}\oplus\mathfrak{H}_3$ algebra}
\label{app_JacobiWittH3}

In this appendix, we develop the analysis of the Jacobi identities and associated constraints for the ansatz \eqref{eq:virh3gendefwithcentral}.

First of all, we consider infinitesimal deformations such that we only keep linear order in the functions. Besides, we can use the fact that $\mathfrak{witt}$ is a rigid subalgebra \cite{fialowski2012formal}, which allows us to set $A_{1}(m,n)=0$ in \eqref{eq:virh3gendefwithcentral}. The Jacobi identities $[\mathcal{L}_{m},[\mathcal{L}_{n},\mathcal{L}_{l}]]+\text{cyclic permutations}=0$ give us three linear relations
\begin{multline}\label{Vir-central}
    (l - m) D_1(l + m, n) + (-l + n) D_1(l + n, m) + (m - n) D_1(m + n, l)=0 \ ,
\end{multline}
\begin{multline}
    [(l - m) C_1(l + m, n) + (-l + n) C_1(l + n, m) + (m - n) C_1(m + n, l)]\mathcal{P}_{l + m + n}=0 \ ,
\end{multline}
and
\begin{multline}
    [(l - m) B_1(l + m, n) + (-l + n) B_1(l + n, m) + (m - n) B_1(m + n, l)]\mathcal{J}_{l + m + n}=0 \ .
\end{multline}

These constraints lead to the following possibilities $D_1(m,n)=(m-n)\tilde{D}_1(m+n)$, $C_1(m,n)=(m-n)\tilde{C}_1(m+n)$ and $B_1(m,n)=(m-n)\tilde{B}_1(m+n)$. The central extension analysis of the Virasoro algebra showed that the solution of \eqref{Vir-central} is $D_1(m,n)=\sigma(m^3-m)\delta_{m+n,0}$.  

Next, the Jacobi identities $[\mathcal{L}_{m},[\mathcal{L}_{n},\mathcal{J}_{l}]]+\text{cyclic permutations}=0$ yield
\begin{multline}
    (m-n) B_2(m+n,l) \mathcal{J}_{l + m + n} +(m-n) C_2(m+n,l) \mathcal{P}_{l + m + n}-lC_{1}(m,n)\delta_{l+m+n,0} + (m-n) D_2(m+n,l) \\ +[-(l + m - n) A_2(m, l) + (l - m + n) A_2(n, l) + (m - n) A_2(m + n, l) ] \mathcal{L}_{l + m + n} = 0 \ ,
\end{multline}
which leads to $B_2(m,n)=C_2(m,n)=0$ and $A_{2}=\alpha(m-n)$. In addition, we find that 
\begin{equation}
    -lC_{1}(m,n)\delta_{l+m+n,0} + (m-n) D_2(m+n,l)=0,
\end{equation}
which is equivalent to $D_2(m+n,l)\overset{!}{=}l\tilde{C}_{1}(m+n)\delta_{l+m+n,0}$.

The Jacobi identities $[\mathcal{L}_{m},[\mathcal{L}_{n},\mathcal{P}_{l}]]+\text{cyclic permutations}=0$ give
\begin{multline}
    (m-n) B_3(m + n, l) \mathcal{J}_{l + m + n} +(m-n) C_3(m + n, l) \mathcal{P}_{l + m + n} -lB_{1}(m,n)\delta_{l+m+n,0} +  (m-n) D_3(m+n, l) \\ +[-(l + m - n) A_3(m, l) + (l - m + n) A_3(n, l) + (m - n) A_3(m + n, l) ] \mathcal{L}_{l + m + n}= 0 \ ,
\end{multline}
which implies $B_3(m,n)=C_3(m,n)=0$ and $A_{3}=\beta(m-n)$. In addition, we find that 
\begin{equation}
    -lB_{1}(m,n)\delta_{l+m+n,0} + (m-n) D_3(m+n,l)=0,
\end{equation}
which is equivalent to $D_3(m+n,l)\overset{!}{=}l\tilde{B}_{1}(m+n)\delta_{l+m+n,0}$.

Similarly, the Jacobi identities $[\mathcal{L}_{m},[\mathcal{P}_{n},\mathcal{P}_{l}]]+\text{cyclic permutations}=0$ and $[\mathcal{L}_{m},[\mathcal{J}_{n},\mathcal{J}_{l}]]+\text{cyclic permutations}=0$  force $A_{4}(m,n)=A_{5}(m,n)=0$. Furthermore, the Jacobi identities $[\mathcal{L}_{m},[\mathcal{J}_{n},\mathcal{P}_{l}]]+\text{cyclic permutations}=0$ cause $A_{6}(m,n)=0$. It follows, as a consequence, that the Jacobi identities for the $\mathfrak{H}_{3}$ sector decouple from $\mathcal{L}$ and are exactly the same as in section \ref{sec_defsHeisenbeg}.

Now, we proceed to explore whether the new infinitesimal deformations we have found are indeed formal or not. For this, we explore the non-linear relations coming from the Jacobi identities.

The Jacobi identities $[\mathcal{L}_{m},[\mathcal{L}_{n},\mathcal{L}_{l}]]+\text{cyclic permutations}=0$ give us 
\begin{equation}
\begin{split}
   & [(m-n-l)(n-l)( \beta \tilde{C_1}(n,l)+\alpha \tilde{B_{1}}(n,l))+(n-m-l)(l-m)(\beta \tilde{C_1}(l,m)+\alpha \tilde{B_{1}}(l,m))\\
    &+(l-m-n)(m-n)(\beta \tilde{C_1}(m,n)+\alpha \tilde{B_{1}}(m,n))]\mathcal{L}_{m+n+l}=0 \ ,
    \end{split}
\end{equation}
and 
\begin{equation}
\begin{split}
  &(n-l)\tilde{C_1}(n,l)D_{3}(m,l+n)+(l-m)\tilde{C_1}(l,m)D_{3}(n,l+m)+(m-n)\tilde{C_1}(m,n)D_{3}(l,m+n)\\
  &+ (n-l)\tilde{B_{1}}(n,l)D_{2}(m,l+n)+(l-m)\tilde{B_{1}}(l,m)D_{2}(n,l+m)+(m-n)\tilde{B_{1}}(m,n)D_{2}(l,m+n)=0 \ .
\end{split}
\end{equation}

Next, the Jacobi identities $[\mathcal{L}_{m},[\mathcal{L}_{n},\mathcal{J}_{l}]]+\text{cyclic permutations}=0$ yield
\begin{equation}
\begin{split}
    [\alpha (l - m) (l + m - n) \tilde{B}_1(l + m, n) - 
   \alpha (l - n) (l - m + n) \tilde{B}_1(l + n, 
     m) + \\ 
     +(m - n) ((-l + m + n) \tilde{B}_1(m, n) F(m + n, l) - 
      \tilde{C}_1(m, n) \bar{F}(l, m + n))] \mathcal{J}_{l + m + n}=0 \ ,
\end{split}      
\end{equation}

\begin{equation}
\begin{split}
    [\alpha (l - m) (l + m - n) \tilde{C}_1(l + m, n) - 
   \alpha (l - n) (l - m + n) \tilde{C}_1(l + n, 
     m) \\
     + (m - n) ((-l + m + n) \tilde{B}_1(m, n) G(m + n, l) - 
      \tilde{C}_1(m, n) \bar{G}(l, m + n))] \mathcal{P}_{l + m + n}=0 \ ,
\end{split}      
\end{equation}
and
\begin{equation}
\begin{split}
    (m - n) (-l + m + n) A(m + n, l) \tilde{B}_1(m, n) + (-m + n) \bar{A}(l, m + n) \tilde{C}_1(
   m, n) + \\
   +\alpha [(l-m)(l+m-n) \tilde{D}_1(l + m, n)-(l-n)(l+n-m) \tilde{D}_1(l + n, m)]=0 \ .
\end{split}      
\end{equation}

The Jacobi identities $[\mathcal{L}_{m},[\mathcal{L}_{n},\mathcal{P}_{l}]]+\text{cyclic permutations}=0$ give
\begin{equation}
\begin{split}
    [\beta (l - m) (l + m - n) \tilde{B}_1(l + m, n) - 
   \beta (l - n) (l - m + n) \tilde{B}_1(l + n, 
     m) + \\ 
     +(m - n) ((-l + m + n) \tilde{C}_1(m, n) \tilde{F}(m + n, l) + 
      \tilde{B}_1(m, n) \bar{F}(l, m + n))] \mathcal{J}_{l + m + n}=0 \ ,
\end{split}      
\end{equation}

\begin{equation}
\begin{split}
    [\beta (l - m) (l + m - n) \tilde{C}_1(l + m, n) - 
   \beta (l - n) (l - m + n) \tilde{C}_1(l + n, 
     m) \\
     + (m - n) ((-l + m + n) \tilde{C}_1(m, n) \tilde{G}(m + n, l) + 
      \tilde{B}_1(m, n) \bar{G}(l, m + n))] \mathcal{P}_{l + m + n}=0 \ ,
\end{split}      
\end{equation}
and
\begin{equation}
\begin{split}
    (m - n) \bar{A}(m + n, l) \tilde{B}_1(m, n) + (m-n)(-l+m+n) \tilde{A}(m+n, l) \tilde{C}_1(
   m, n) + \\
   +\beta [(l-m)(l+m-n) \tilde{D}_1(l + m, n)-(l-n)(l+n-m) \tilde{D}_1(l + n, m)]=0 \ .
\end{split}      
\end{equation}

The Jacobi identities $[\mathcal{L}_{m},[\mathcal{J}_{n},\mathcal{J}_{l}]]+\text{cyclic permutations}=0$ yield
\begin{equation}
\alpha (l-m) D_2(l+m,n) + 
 \alpha (m-n) D_2(m+n, 
   l) + (l-n) [D_2(m,l+n) F(n,l) + D_3(m,l+n) G(n,l)]=0 \ ,
\end{equation}
and
\begin{equation}
     (l-n)(l-m+n)[\alpha^2-\alpha F(n,l)-\beta G(n,l)] \mathcal{L}_{l + m + n} = 0  \ . 
\end{equation}

The Jacobi identities $[\mathcal{L}_{m},[\mathcal{P}_{n},\mathcal{P}_{l}]]+\text{cyclic permutations}=0$ give rise to
\begin{equation}
\beta (l-m) D_3(l+m,n)+\beta (m-n) D_3(m+n,l) + (l-n)[D_2(m,l+n) \tilde{F}(n, l) + D_3(m,l+n) \tilde{G}(n,l)]=0 \ ,
\end{equation}
and
\begin{equation}
    (l-n)(l-m+n)[\beta^2-\beta \tilde{G}(n,l)-\alpha \tilde{F}(n,l)] \mathcal{L}_{l + m + n} = 0  \ . 
\end{equation}

Finally, the Jacobi identities $[\mathcal{L}_{m},[\mathcal{J}_{n},\mathcal{P}_{l}]]+\text{cyclic permutations}=0$ lead to
\begin{equation}
\begin{split}
  &\bar{F}(n,l)D_{2}(m,l+n)+\bar{G}(n,l)D_{3}(m,l+n)+\alpha(n-m)D_{3}(m+n,l)+\beta(m-l)D_{2}(l+m,n)=0 \ ,
\end{split}
\end{equation}
and
\begin{equation}
     (l-m+n)[\alpha\beta (l-n)+\beta \bar{G}(n,l)+\alpha \bar{F}(n,l)] \mathcal{L}_{l + m + n} = 0 \ . 
\end{equation}

\section{Heisenberg from contractions}
\label{app_Heisenbergfromcontractions}


\subsection{The most general contractions of two Virasoro algebras}

We are going to investigate the most general contractions of two Virasoro algebras.
\begin{align}\label{twovir-contract1}
 & [\mathcal{L}_{m},\mathcal{L}_{n}]=\frac{1}{\varrho}(m-n){\mathcal{L}}_{m+n}+\frac{c}{12\varrho^2}(m^3-\alpha m)\delta_{n+m,0} \ , \cr
 &[\mathcal{L}_{m},\bar{\mathcal{L}}_{n}]=0 \ ,\\
 &[\bar{\mathcal{L}}_{m},\bar{\mathcal{L}}_{n}]=\frac{1}{\vartheta}(m-n)\bar{\mathcal{L}}_{m+n}+\frac{\bar{c}}{12\vartheta^2}(m^3-{\beta m})\delta_{n+m,0} \ ,\nonumber
\end{align} 
in which $\varrho$ and $\vartheta$ are contraction parameters. The various limits of these parameters lead to different algebras which are classified as follows:

\begin{enumerate}
    \item $\vartheta\rightarrow \infty$ which leads to take the direct sum of one Virasoro with an Abelian ideal algebra as 
    \begin{align}
 & [\mathcal{L}_{m},\mathcal{L}_{n}]=(m-n){\mathcal{L}}_{m+n}+\frac{c}{12}(m^3-\alpha m)\delta_{n+m,0} \ , \cr
 &[\mathcal{L}_{m},\bar{\mathcal{L}}_{n}]=0 \ ,\\
 &[\bar{\mathcal{L}}_{m},\bar{\mathcal{L}}_{n}]=0 \ .\nonumber
\end{align} 

\item The next case is when we take the limit $\vartheta\rightarrow \infty$ while keeping $\frac{\bar{c}\beta}{12\vartheta^2}$ to be finite. This leads to the direct sum of a Virasoro algebra with a current algebra as 
\begin{align}
 & [\mathcal{L}_{m},\mathcal{L}_{n}]=(m-n){\mathcal{L}}_{m+n}+\frac{c}{12\varrho}(m^3-\alpha m)\delta_{n+m,0} \ , \cr
 &[\mathcal{L}_{m},\bar{\mathcal{L}}_{n}]=0 \ ,\\
 &[\bar{\mathcal{L}}_{m},\bar{\mathcal{L}}_{n}]=\frac{\bar{c}\beta }{12\vartheta}m\delta_{n+m,0} \ ,\nonumber
\end{align} 

\paragraph{Remark.} One can also consider the algebra \eqref{twovir-contract1} without linear central term and get the same result. To this end, one starts with the Virasoro algebra
\be
[\bar{\mathcal{L}}_{m},\bar{\mathcal{L}}_{n}]=(m-n)\bar{\mathcal{L}}_{m+n}+\frac{\bar{c}\beta }{12\vartheta}m^3\delta_{n+m,0} \ ,
\ee
and uses the redefinition as
\be
{\cal J}_m :=\frac1{m\varepsilon} \bar{\mathcal{L}}_{m}, \quad \varepsilon\to\infty, \qquad \frac{c}{12 \varepsilon^2}=\hbar=\text{fixed} \ .
\ee
It is then immediate to see that in the limit $\varepsilon\to \infty$ and keeping ${\cal J}_n, \text{with} \ \hbar$ fixed, we obtain
\be\label{JJ-hbar}
[\mathcal{J}_{m},\mathcal{J}_{n}]={\hbar}\ n\delta_{m+n,0} \ . 
\ee

\item The limit $\vartheta\rightarrow \infty$ and $\varrho\rightarrow \infty$ takes us to the direct sum of two current algebras as
\begin{align}\label{twocurrents-diffc}
 & [\mathcal{L}_{m},\mathcal{L}_{n}]=\hbar\, m\,\delta_{n+m,0} \ , \cr
 &[\mathcal{L}_{m},\bar{\mathcal{L}}_{n}]=0 \ ,\\
 &[\bar{\mathcal{L}}_{m},\bar{\mathcal{L}}_{n}]=\bar{\hbar}\,m\,\delta_{n+m,0} \ ,\nonumber
\end{align} 
where two coefficients $\hbar=\frac{c\alpha}{12\varrho^2}$ and $\bar{\hbar}=\frac{\bar{c}\beta }{12\vartheta^2}$ are kept to be finite.
As we mentioned before, this algebra is isomorphic to \eqref{Dirac-3d} (by a complex redefinition), which can be also obtained as another contraction of two Virasoro algebras in a different basis (see Appendix \ref{appendixB2}). 

\item The limit $\vartheta\rightarrow \infty$ and $\varrho\rightarrow \infty$ takes us to the direct sum of a current algebra and an Abelian ideal algebra as
\begin{align}
 & [\mathcal{L}_{m},\mathcal{L}_{n}]=\hbar\,m\,\delta_{n+m,0} \ , \cr
 &[\mathcal{L}_{m},\bar{\mathcal{L}}_{n}]=0 \ ,\\
 &[\bar{\mathcal{L}}_{m},\bar{\mathcal{L}}_{n}]=0 \ ,\nonumber
\end{align} 
where the coefficient $\hbar=\frac{c\alpha}{12\varrho^2}$ is kept to be finite.

\item The limit $\vartheta\rightarrow \infty$ and $\varrho\rightarrow \infty$ yields a trivial contraction in which all the generators commute with each other. 
\end{enumerate}

Next, we explore contractions of two Virasoro algebras in a new basis as
\begin{align}\label{twoVir-contract}
 & [\mathcal{J}_{m},\mathcal{J}_{n}]=\frac{1}{\varepsilon}(m-n)\mathcal{J}_{m+n}+\frac{c_{JJ}}{12\varepsilon^2}(m^3-{\alpha m})\delta_{n+m,0} \ , \cr
 &[\mathcal{J}_{m},\mathcal{P}_{n}]=\frac{1}{\varepsilon}(m-n)\mathcal{P}_{m+n}+\frac{c_{JP}}{12\varepsilon\varsigma}(m^3-{\bar{\alpha} m})\delta_{n+m,0} \ ,\\
 &[\mathcal{P}_{m},\mathcal{P}_{n}]=\frac{\varepsilon}{\varsigma^2}(m-n)\mathcal{J}_{m+n}+\frac{c_{JJ}}{12\varsigma^2}(m^3-{{\alpha} m})\delta_{n+m,0} \ , \nonumber
\end{align}
where $\varepsilon$ and $\varsigma$ are new contractions parameters. Also, one easily checks that $c_{JJ}=c-\bar{c}$ and $c_{JP}=c+\bar{c}$. One can obtain \eqref{twoVir-contract} from \eqref{twovir-contract1} by using the redefinition of generators as 
\begin{equation}\label{Vir-BMS}
\mathcal{L}_m=\frac12(\mathcal{J}_m+{\mathcal{P}_m}).\ \ \ \bar{\mathcal{L}}_{-m}=\frac12(\mathcal{P}_m-{\mathcal{J}_m}),
\end{equation}

Now we consider different limits of contraction parameters $\varepsilon$ and $\varsigma$:

\begin{enumerate}
    \item $\varsigma\rightarrow\infty$ while keeping $\varepsilon$ to be finite. 
    For this case, one can obtain an algebra as 
    \begin{align}\label{Vir-contraction}
 & [\mathcal{J}_{m},\mathcal{J}_{n}]=(m-n)\mathcal{J}_{m+n}+\frac{c_{JJ}}{12}(m^3-{\alpha m})\delta_{n+m,0} \ , \cr
 &[\mathcal{J}_{m},\mathcal{P}_{n}]=(m-n)\mathcal{P}_{m+n}+\frac{\bar{c}_{JP}}{12}(m^3-{\bar{\alpha} m})\delta_{n+m,0} \ ,\\
 &[\mathcal{P}_{m},\mathcal{P}_{n}]=0 \ ,\nonumber
\end{align}
which is nothing but the central extension of $\mathfrak{bms}_{3}$ algebra. Notice that we kept $\frac{\bar{c}_{JP}}{12}=\frac{c_{JP}}{12\varsigma}$ to be also finite. 

\item $\varepsilon\rightarrow\infty$ as well as $\varsigma^2 \rightarrow\infty$ while keeping $\frac{\varepsilon}{\varsigma^2}=\nu$ to be finite. The final algebra takes the form
\begin{align}\label{Vir-contraction-1}
 & [\mathcal{J}_{m},\mathcal{J}_{n}]=0 \ , \cr
 &[\mathcal{J}_{m},\mathcal{P}_{n}]=\hbar{m}\delta_{n+m,0} \ ,\\
 &[\mathcal{P}_{m},\mathcal{P}_{n}]=\nu(m-n)\mathcal{J}_{m+n} \ ,\nonumber
\end{align}
in which we assumed that $\hbar=\frac{\bar{\alpha} c_{JP}}{12\varepsilon\varsigma}$ is finite. 
The algebra \eqref{Vir-contraction-1}, which is obtained by contraction of two copies of Virasoro, is isomorphic to the algebras \eqref{MostDeform-JJ} and \eqref{Deform-PP-1}.

\item $\varepsilon\rightarrow\infty$ as well as $\varsigma^2 \rightarrow\infty$ while keeping $\hbar=\frac{\bar{\alpha} c_{JP}}{12\varepsilon\varsigma}$ to be finite. The final algebra takes the form
\begin{align}\label{Vir-contraction-2}
 & [\mathcal{J}_{m},\mathcal{J}_{n}]=0 \ , \cr
 &[\mathcal{J}_{m},\mathcal{P}_{n}]=\hbar\,{m}\,\delta_{n+m,0} \ ,\\
 &[\mathcal{P}_{m},\mathcal{P}_{n}]=0 \ ,\nonumber
\end{align}
which is exactly the same as Heisenberg algebra \eqref{Dirac-3d}, whose deformations we considered in this work.

\begin{itemize}
    \item There are two other options which can be obtained as contraction of two Virasoro algebras in this basis:
    \begin{align}
 & [\mathcal{J}_{m},\mathcal{J}_{n}]=\bar{\hbar}\,m\,\delta_{m+n,0} \ , \cr
 &[\mathcal{J}_{m},\mathcal{P}_{n}]=\hbar\,{m}\,\delta_{n+m,0} \ ,\\
 &[\mathcal{P}_{m},\mathcal{P}_{n}]=0 \ ,\nonumber
\end{align}

where $\bar{\hbar}=\frac{c_{JJ}\alpha }{12\varepsilon^2}$, and 
\begin{align}
 & [\mathcal{J}_{m},\mathcal{J}_{n}]=\bar{\hbar}\,m\,\delta_{m+n,0} \ , \cr
 &[\mathcal{J}_{m},\mathcal{P}_{n}]=\hbar\,{m}\,\delta_{n+m,0} \ ,\\
 &[\mathcal{P}_{m},\mathcal{P}_{n}]=\tilde{\hbar}\,m\,\delta_{m+n,0} \ ,\nonumber
\end{align}
\end{itemize}
where $\tilde{\hbar}=\frac{c_{JJ}\alpha}{12\varsigma^2}$. Then, it can be shown that, by an appropriate redefinition of generators, both of them are equivalent to \eqref{Vir-contraction-2}, as long as $\tilde{\hbar}$ is related to $\bar{\hbar}$. 
\end{enumerate}

\subsection{Connection of $\mathfrak{vir}\oplus\mathfrak{vir}$ with $\mathfrak{H}_3$ - twisted Sugawara construction}\label{appendixB2}
 
The above contraction constitutes the inverse procedure of the twisted Sugawara construction, the ``twisted Sugawara contraction''. To see it, let us start with 
\be\label{T-Sug}
\mathcal{L}_{n}=\frac{1}{2\hbar}\sum_m \mathcal{J}_{n-m}\mathcal{J}_{m}+ \varepsilon n \mathcal{J}_{n} \ .
\ee
Clearly, the $\mathcal{L}_{n}$ satisfy the Virasoro algebra with central charge $\frac{c}{12}=\varepsilon^2\hbar$. In the limit $\varepsilon\to\infty,\ \hbar=fixed$, \eqref{T-Sug} clearly shows how $\mathcal{L}_n$ can be reduced to $\mathcal{J}_n$.

We close this part with a comment. The current algebra \eqref{JJ-hbar} is closely related to the Heisenberg algebra. Let 
\be\label{Current-to-Heisenberg}
X_n:=\frac{1}{\sqrt{2n}}(\mathcal{J}_n+i \mathcal{J}_{-n}) \ ,\qquad P_n:=\frac{i}{\sqrt{2n}}(\mathcal{J}_{-n}+i \mathcal{J}_{n}) \ , \quad n>0 \ . 
\ee
It is then immediate to realize that $[X_n, P_m]=i\hbar \delta_{m,n}$. Similarly, starting with the direct sum of two current algebras
\begin{align}\label{JJ+--hbar}
&[\mathcal{J}^{+}_{m},\mathcal{J}^{+}_{n}]={\hbar}\ n \delta_{m+n,0} \ , \nonumber\\
&[\mathcal{J}^{-}_{m},\mathcal{J}^{-}_{n}]={\hbar}\ n \delta_{m+n,0} \ , \\
&[\mathcal{J}^{+}_{m},\mathcal{J}^{-}_{n}]=0 \ , \nonumber
\end{align}
one may construct
\be\label{Current-to-Heisenberg1}
X_n:=\left\{\begin{array}{cc} \frac{1}{\sqrt{2n}}(\mathcal{J}^+_n+i \mathcal{J}^+_{-n}) \qquad n>0 \\ \frac{1}{\sqrt{-2n}}(\mathcal{J}^-_n+i \mathcal{J}^-_{-n}) \qquad n<0 \ ,
\end{array}\right.\qquad P_n:=\left\{\begin{array}{cc} \frac{i}{\sqrt{2n}}(\mathcal{J}^+_{-n}+i \mathcal{J}^+_{n}) \qquad n>0 \\ \frac{i}{\sqrt{-2n}}(\mathcal{J}^-_n+i \mathcal{J}^-_{-n}) \qquad n<0 \ ,\end{array}\right.
\ee
such that $[X_n,P_m]=i\hbar \delta_{m,n}$. We note, however, that $\mathcal{J}^\pm_0$, which commute with all the other generators, do not enter the Heisenberg algebra of $X_n, P_m$ and also that $X_0, P_0$ are not defined. One may add the latter to the algebra by hand.


\bibliographystyle{fullsort.bst}
 
\bibliography{references}


\end{document}